\documentclass[%
 reprint,
 amsmath,amssymb,
 pre,
]{revtex4-1}
\usepackage{graphicx}
\usepackage{color}
\usepackage{dcolumn}
\usepackage{bm}
\usepackage{hyperref}

\graphicspath{{figs/}}
\definecolor{mygreen}{rgb}{0,0.5,0}
\newcommand{\colg}[1]{\textcolor{mygreen}{#1}}
\newcommand{\colga}[1]{\textcolor{green}{#1}}
\newcommand{\colb}[1]{\textcolor{blue}{#1}}
\newcommand{\colr}[1]{\textcolor{red}{#1}}
\newcommand{\colm}[1]{\textcolor{magenta}{#1}}

\newcommand{\colc}[1]{\textcolor{cyan}{#1}}
\def\var{\textrm{Var}}
\def\beq {\begin{equation}}
\def\eeq {\end{equation}}
\def\beqa {\begin{eqnarray}}
\def\eeqa {\end{eqnarray}}
\def \bnum {\begin{enumerate}}
\def \enum {\end{enumerate}}
\def\bi {\begin{itemize}}
\def\ei {\end{itemize}}
\def \bdes {\begin{description}}
\def \edes {\end{description}}
\def\meandiss {\langle\epsilon\rangle}
\def\rel {R_{\lambda}}
\def\rer {Re_{r}}

\def\app{\approx}

\def\dho{\partial}
\def\dhot{\partial t}

\def\epsrvec{\epsilon_r(\textbf{x},t)}

\def\epsr{\epsilon_r}
\def\epsrmn{\langle{\epsilon_r}\rangle}
\def\epsroned{\hat{\epsilon}_r}
\def\normr {r/\eta}
\def\keta {k\eta}
\def\koneeta {k_1\eta}

\def\la {\langle}
\def\ra {\rangle}

\def\mbf {\mathbf}
\def\kmax {k_{max}}

\def\delx{\Delta x}

\def\kres{{k_{max}\eta}}

\def\ur { \mathbf{r} }

\def\ur { \mathbf{r} }

\def\uu { \mathbf{u} }
\def\psip { \varsigma_p }
\def\taup { \tau_{p/3} }

\def\ux { \mathbf{x} }

\def\epsr{{\epsilon_r}}
\def\repsr{{r\epsilon_r}}
\def\repsrpow{{(r\epsilon_r)^{1/3}}}
\def\delur{{\Delta u(r)}}

\def\modelur{{|\Delta u(r)|}}
\def\modv{{|V|}}
\def\bigd{{\big|}}

\def\eps{{\epsilon}}

\begin{document}

\preprint{APS/123-QED}

\title{Refined similarity hypothesis using three-dimensional local averages}

\author{Kartik P. Iyer}
\email{kartik.iyer@roma2.infn.it}
\affiliation{
Department of Physics and INFN, 
University of Rome Tor Vergata,
Rome, 00133, Italy
}
\author{Katepalli R. Sreenivasan}
\affiliation{
Departments of Physics and Mechanical Engineering and the Courant Institute of Mathematical Sciences \\
New York University,
Brooklyn, 11201, USA
}
\author{P.K. Yeung}
\affiliation{
Schools of Aerospace Engineering and Mechanical Engineering,
Georgia Institute of Technology,
Atlanta, 30332, USA
}
\date{\textbf{Postprint version of the manuscript published in Phys. Rev. E,
$92$, $063024$ $(2015)$}}

\begin{abstract}
The refined similarity hypotheses of Kolmogorov, regarded as an important ingredient of intermittent turbulence, 
has been tested in the past using one-dimensional data and plausible surrogates of energy dissipation. 
We employ data from direct numerical simulations, 
at the microscale Reynolds number $\rel \sim 650$, 
on a periodic box of $4096^3$ grid points to test the hypotheses using 
3D averages. 
In particular, we study the small-scale properties of the stochastic 
variable $V = \delur/\repsrpow$, 
where $\delur$ is the longitudinal velocity increment and 
$\epsr$ is the dissipation rate averaged over a 
three-dimensional volume of linear size $r$. 
We show that $V$ is universal in the inertial subrange.
In the dissipation range, 
the statistics of $V$ are shown to depend 
solely on a local Reynolds number.
\begin{description}
\item[PACS numbers]
May be entered using the \verb+\pacs{#1}+ command.
\end{description}
\end{abstract}

\pacs{Valid PACS appear here}
\maketitle

\section{\label{sec:level1}Introduction}
A statistical description of the local flow structure in high-Reynolds number turbulence, known as K41, was given in 
Ref.~[\onlinecite{K41a}]. 
In its simplified version, the first hypothesis relates the probability density function (PDF) of the longitudinal velocity increments $\Delta u(r) = [\uu(\ux+\ur)-\uu(\ux)].\ur/|\ur|$ to the mean energy dissipation rate $\meandiss$ and the fluid viscosity ($\nu$), for spatial separations $r=|\ur| \ll L$,  $L$ being the integral scale of the turbulence. The second hypothesis is that if the Reynolds number is very large, there exists a range of scales (the so-called inertial range) for which $\nu$ becomes irrelevant, so that the PDF of $\Delta u(r)$ depends only on $\meandiss$, apart from $r$ itself.  An exact result for the third moment of $\delur$ in the inertial range is given in 
Ref.~[\onlinecite{K41b}] as
\beq
\label{k41.eq}
\la (\delur)^3 \ra = -\frac{4}{5}\meandiss r.
\eeq 
An implicit assumption in K41 is that the rate of transfer of energy from the large to the small scales is a constant (or mildly varying) everywhere in the flow and is equal to (or not far from) $\meandiss$. However, the energy dissipation rate per unit mass of a turbulent fluid, given by
\beq
\eps(\ux,t) = \frac{\nu}{2} \sum_{i,j} \big(\frac{\dho u_i}{\dho x_j} + 
\frac{\dho u_j}{\dho x_i}\big )^2\;,
\eeq
fluctuates wildly in space and time[\onlinecite{YDS2012}]. The fluctuations of $\eps (\ux,t)$ may depend on the Reynolds number and the large-scale properties, which can be non-universal.
(Refs.~[\onlinecite{MY.II,Fri95,SA97}]).

The refined similarity theory 
(Ref.~[\onlinecite{K62}]), 
also known as K62, 
introduced more restrictive alternatives and abandoned strict universality. It postulated that one of the most important factors determining the statistics of $\Delta u(r)$ for $r \ll L$
is the dissipation rate averaged over a local volume $\mathcal{V}(r)$ of linear dimension $r$,
i.e., the quantity (Ref.~[\onlinecite{obukhov62}])
\beq
\label{3d_dissavg.eq}
\epsrvec = \frac{1}{r^3}\int_{\mathcal{V}(r)}
\epsilon(\textbf{x}+\textbf{r}',t) d\textbf{r}' \;.
\eeq
Following Ref.~[\onlinecite{K62}], 
the quantities $r$ and $\epsr(\ux,t)$ can be used to construct a velocity scale at the point $(\ux,t)$ as
$U_r = (r\epsr)^{1/3}$, and a local Reynolds number can be formed as
\beq
\label{locre.eq}
\rer = \frac{U_r r}{\nu} = \frac{(r\epsr)^{1/3}r}{\nu} = 
\big (\frac{r}{\eta_r} \big )^{4/3},\qquad 
\eta_r = \big (\frac{\nu^3}{\epsr}\big )^{1/4}\;,
\eeq
where $\eta_r$ is the local Kolmogorov length scale.  The first refined similarity hypothesis can be stated
(Ref.~[\onlinecite{krsh}]),
for $r \ll L$, as
\beq
\label{k621.eq}
\Delta u(r) \equiv V\repsrpow\;,
\eeq
where $V$ is a dimensionless
stochastic variable whose PDF depends only on $\rer$.
The second refined hypothesis states that if $\rer \gg 1$, 
the PDF of $V$ becomes independent of $\rer$, i.e., it is universal.

The first and the second hypotheses relate the scaling
exponents of the local dissipation to that of the velocity 
increments \cite{Fri95}.
In particular, the exponent $\psip$ of
velocity increments is related to the exponent $\tau_p$ of the
the local dissipation for the moment of order $p$ as
\beq
\label{mfrac.eq}
\psip = \frac{p}{3}+\taup \;.
\eeq
By independently measuring the scaling exponents of velocity and local 
dissipation at the small scales, Eq.~\ref{mfrac.eq} can be used
to verify the K62 hypotheses \cite{huang14}. 
However, such an analysis of Eq.~\ref{k621.eq}, 
can obscure the importance of the 
stochastic function $V$. For instance,
the third moment of $V$ is linked to the mechanism of vortex stretching
and energy transfer between different scales [\onlinecite{krsh}]. 
Moreover, although $V$ is the ratio of two intermittent quantities, it is itself
not intermittent [\onlinecite{krsh}] and is less sensitive
to finite sampling effects.

Previous work 
(Refs.~[\onlinecite{krsh}],[\onlinecite{wangkrsh}]) 
have examined the two refined similarity hypotheses using one-dimensional (1D) averages of the local dissipation rate $\hat{\epsr}(\ux,t)$, calculated over a line $\mathcal{L}$ of length $r$ as,
\beq
\label{1diss.eq}
\hat{\epsilon}_r(\ux,t) = \frac{1}{r} \int_\mathcal{L} \epsilon(\ux+\ur',t) dr'.
\eeq
The use of 1D averages over 3D averages in experiments is partly forced by difficulties in obtaining well-resolved measurements in 3D at high-Reynolds-numbers, especially for small spatial separations $r \ll L$. Computationally, 1D averages are easier to calculate than 3D averages, since the latter may require heavy communication between different processors in a parallel network, especially when the spatial separation $r$ is large.

In this paper,  we examine the first and the second postulates of the refined similarity theory using 3D local averages of dissipation.
A brief overview of the numerical procedure is given in section II. Some properties of 3D averaged dissipation are juxtaposed with those of 1D averaged dissipation in section III.  Results for the K62 theory are presented for a $4096^3$ simulation at $\rel \sim 650$ in section IV. We summarize the main results in section V.
\section{Numerical Procedure}
\subsection{Direct Numerical Simulations}
The fluctuating velocity field $\uu(\ux,t)$ obeys the forced Navier-Stokes equations for
an incompressible flow as
\beqa
\nabla \cdot \uu &=& 0 \; ,\\
\dho\uu/\dhot  + \uu\cdot\nabla{\uu}
&=&
-{\nabla} p/\rho  +  \nu {\nabla}^{2}\uu  + \mbf{f} \;,
\eeqa 
where, $\rho$ is the constant fluid density and $\mbf{f}$ is the random
forcing used to achieve stationarity 
[\onlinecite{EP88}]. 
The domain is a
periodic cube with edge length
$2\pi$ with $N$ grid points to a side. 
A Fourier pseudospectral method is used for spatial
discretization and the equations are integrated in time using a second order
Runge-Kutta scheme as given in 
Ref.~[\onlinecite{rogallo}]. 
Aliasing errors from the nonlinear
term are effectively controlled by removing all coefficients
with wave-number magnitude greater than $\kmax = (\sqrt 2/3)N$.
The non-dimensional parameter $\kres$ gives the resolution of the simulation.
The ratio of the grid spacing ($\delx = 2\pi/N$) to the Kolmogorov length scale ($\eta$) is,
$\delx/\eta \app 2.96/(\kres)$. 
Parameters of interest for the simulations used
in this work are summarized in Table \ref{sim1.tab}.
\begin{table}[b]
\caption{Summary of the microscale Reynolds number ($\rel$), number of grid
points ($N^3$) and
resolution ($\kres$) for the periodic cube with edge length $2\pi$
studied in this work.
}
\begin{ruledtabular}
\begin{tabular}{lcc}
$\rel$ & $N^3$ & $\kres$ \\ \hline \\
240 & $512^3$ & 1.4  \\
240 & $2048^3$ & 5.7 \\
650 & $4096^3$ & 2.7 \\
\end{tabular}
\end{ruledtabular}
\label{sim1.tab}
\end{table}
Details about the parallel implementation are given in 
Ref.~[\onlinecite{DYP2008}].  
\subsection{Local 3D averages}
Samples of the 1D and 3D local averages of dissipation, 
($\epsroned$, $\epsr$)  
are calculated at the point $(x_1,y_1,z_1)$ 
over a cube with edge
length of $r$ grid spacing using the formulae,
\beqa
\label{boxavgoned.eq}
\epsroned(x_1,y_1,z_1) &=& \frac{1}{(r+1)}
\sum_{x=x_1}^{x_{r+1}}\epsilon(x,y_1,z_1)\; , \\
\label{boxavg.eq}
\epsr(x_1,y_1,z_1) &=& \frac{1}{(r+1)^3}{\sum_{z=z_1}^{z_{r+1}}
\sum_{y=y_1}^{y_{r+1}}\sum_{x=x_1}^{x_{r+1}}\epsilon(x,y,z)}\; .
\eeqa
In order to obtain adequately converged statistics of $\epsroned$ and $\epsr$,
Eqs.~\ref{boxavgoned.eq} and \ref{boxavg.eq} need to be used at 
every point $(x,y,z)$ in a $N^3$ grid. The 1D averages are computed along
the three orthogonal directions for adequate sampling.
The examination of the statistics of $\epsroned$ and $\epsr$ at various scale
sizes, requires Eqs.~\ref{boxavgoned.eq} and \ref{boxavg.eq}
 to be used at all 
non-trivial spatial separations, which spans 
$r=1,2,\ldots,N/2$ grid spacings. To save computer time, the calculations
are performed only at selected multiples
of grid spacing, at the larger spatial separations.
In the case of 3D local averages,
added to the computational complexity, is the
inter-processor communication time due to 
the domain decomposition in a parallel algorithm. 
The significant computation and communication costs involved,
makes the  calculation of 
3D local averages
very challenging, especially at higher Reynolds numbers. 
Details of the parallel algorithm we used
to compute the 3D local averages can be found in 
Ref.~[\onlinecite{kp2014}].
\subsection{Calculation of stochastic variable $V$}
Consider the stochastic variable $V$ (Eq.~\ref{k621.eq})
defined as
\beq
\label{v.eq}
V(r) = \delur/\repsrpow \;,
\eeq
where $\delur = [\uu(\ux+\ur)-\uu(\ux)]\cdot \ur/|\ur|$ is
the longitudinal velocity increment along separation vector
$\ur$ , whose magnitude is $r = |\ur|$ and 
$\epsr$ is the 3D local average of
dissipation.

It may be noted here that the probability distribution of $V$ 
can be regarded as a ratio distribution since $V$ is the ratio of random
variables (Eq.~\ref{v.eq}). 
At scales $r \ll L$, both the velocity increments
and the local dissipation $\epsr$ are intermittent and essentially 
non-Gaussian.
All non-negative moments of 
$V(r)$ for $r \ll L$ are well-defined and finite.
In particular, $V$ is not a Cauchy variable [\onlinecite{bio69}]
at least at the small scales ($r \ll L$).

\begin{figure}[!]
\begin{minipage} [t] {0.5\textwidth}
\includegraphics [width=0.6\textwidth]{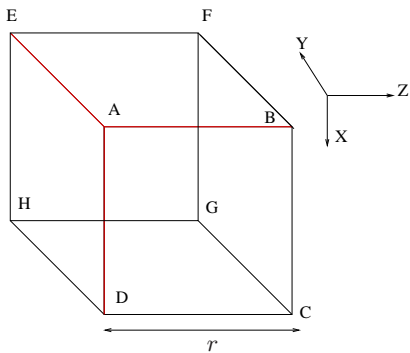}
\begin{picture}(1,1)
\put(-80,-7){$r$}
\end{picture}
\end{minipage}
\caption{(Color online) Cube of edge length $r$ used for the
calculation of $V(r)$.
Line segments AD, AE and AB denote edges along which $V$ 
(Eq.~\ref{valpha.eq})
 is
calculated in the $X$, $Y$ and $Z$ directions respectively.}
\label{box.fig}
\end{figure}

Since a cube such as that in
Fig.~\ref{box.fig} has twelve edges over which a longitudinal
velocity increments can be  defined, we can have twelve different samples
of $V$. Presumably in isotropic homogeneous
 turbulence the statistics
of $V$ along these twelve directions can be considered as
different samples which can then be averaged.
In this work we consider samples of $V(r)$  along three edges
of a
cube of length $r$ as
\beq
\label{valpha.eq}
V_\alpha(r) = \frac{\Delta u_\alpha(r)}{\repsrpow} ,
\quad \; \alpha=1,2,3 \;. 
\eeq
Here, $\Delta u_\alpha (r)$ for $\alpha=1,2,3$
correspond to  the longitudinal
velocity increments along edges AD, AE
and AB respectively in Fig.~\ref{box.fig}.
The samples $V_\alpha$ for
$\alpha=1,2,3$ can be considered as 
different realizations for
the statistics of $V$ and can be averaged accordingly.
The use of three different samples of $V$ in this manner improves the statistical stability, especially when the
averaging length $r$ is small ($r \sim \eta$).
\section{Results}
\subsection{Local averages of dissipation}
\begin{figure}[!]
\begin{minipage}{0.5\textwidth}
\includegraphics [width=0.604\textwidth]{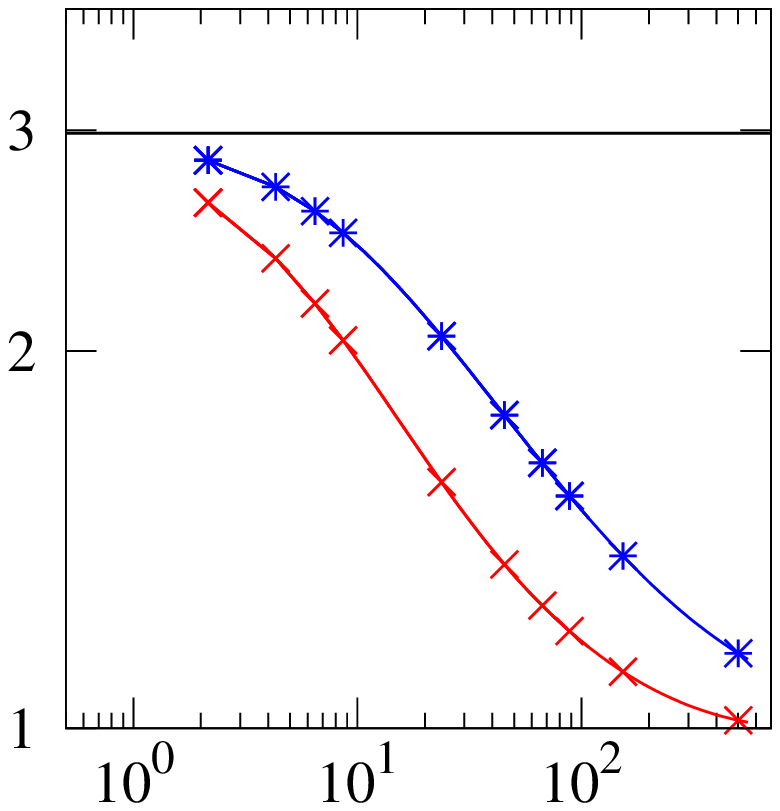}
\hspace{-6.4em}
\includegraphics [width=0.604\textwidth]{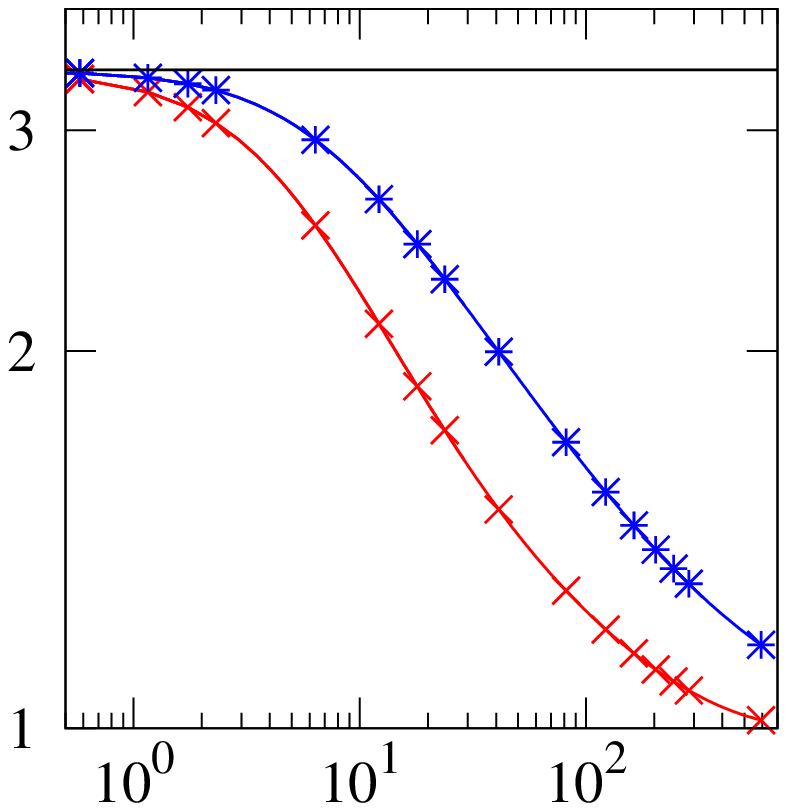}
\begin{picture}(1,1)
\put(45,2){$\normr$}
\put(-55,2){$\normr$}
\put(-80,35){(a)}
\put(20,35){(b)}
\put(-115,20){\rotatebox{90}{$\la\epsr^2\ra/\epsrmn^2,\;\la\epsroned^2\ra/
\la\epsroned\ra^2 $}}
\end{picture}
\end{minipage}
\protect\caption{(Color online) Second order moments of 3D local averages (\colr{$\times$}) and
1D local averages (\colb{$\ast$}) of dissipation for 
(a) $\kres = 1.4$, (b) $\kres = 5.7$, both at $\rel \sim 240$. 
Horizontal solid lines correspond to second order dissipation moment
($\la \epsilon^2\ra/\la \epsilon \ra^2$).
}
\label{secmom.fig}
\end{figure}

Figure \ref{secmom.fig} shows the second moments of local 3D and 1D
averaged dissipation as a function of spatial separation for
two different resolutions at a given Reynolds number. It is
clear that the 3D local average is more resolution limited than
the 1D average for smaller separations. 
Consider local averages
over a cube with edge length $\Delta$.
A sample of the 1D average represents the 
dissipation at the midpoint of an edge in the cube,
whereas the
3D average represents the dissipation at the centroid of the cube,
which is $\sqrt{3}$ times further away from the given grid point
than the corresponding 1D average. Hence 1D averages are
closer to the point-wise averages than their 3D counterparts
and thereby have better spatial resolution.

For a given spatial separation, the second moment of 3D 
averaged dissipation is smaller than
that of 1D averaged dissipation (Fig.~\ref{secmom.fig}), 
indicating that $\epsr$ may be
less intermittent than $\epsroned$ 
(see also [\onlinecite{krsad}]). 
It has been verified (although not shown here) that this is also true 
for all higher order moments.
For a given averaging length $r$,
we can indeed write
\beq
\label{dissord.eq}
\la \epsr^q \ra \le \la \epsroned^q \ra \le \la \epsilon^q \ra \;,
\quad q=1,2,3,\ldots,
\eeq
equality occurring when $q=1$.
 
\begin{figure}
\begin{minipage}{0.5\textwidth}
\includegraphics [width=0.593\textwidth]{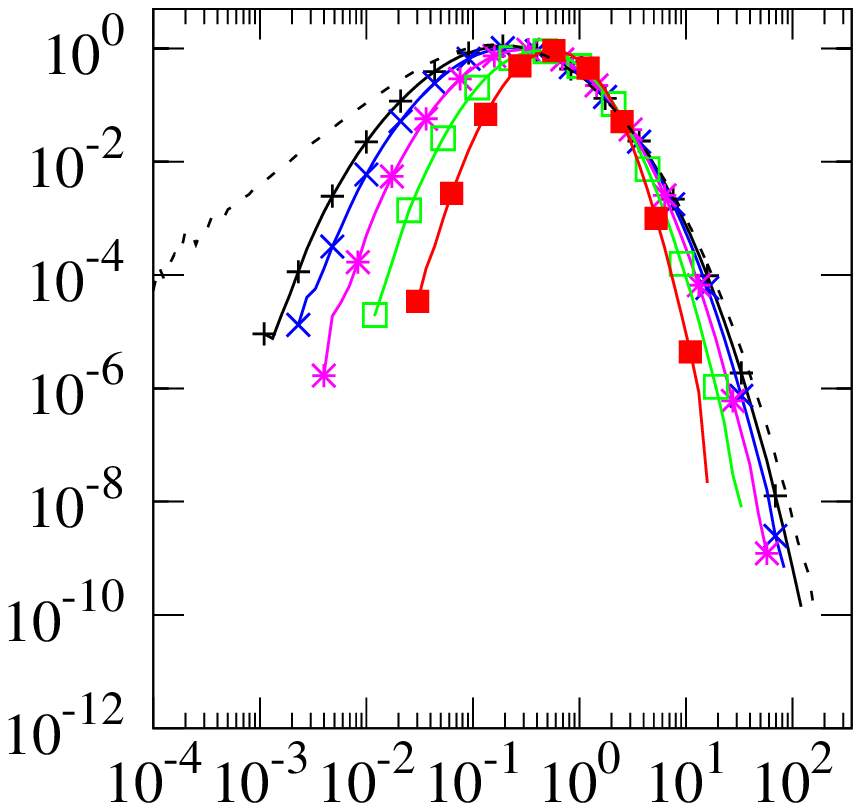}
\hspace{-5.8em}
\includegraphics [width=0.593\textwidth]{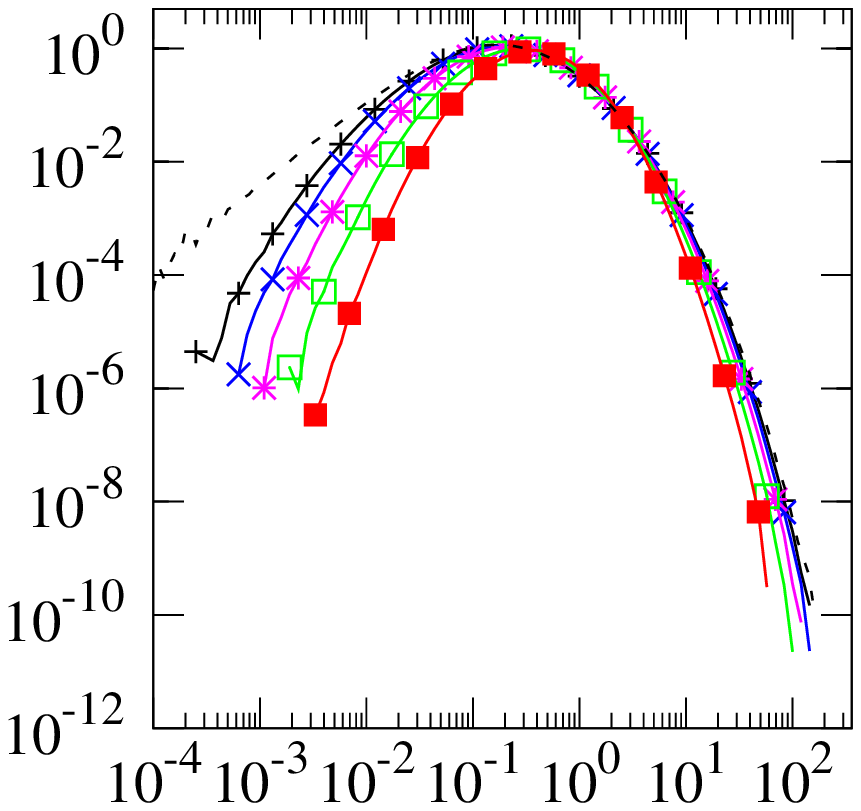} \\
\includegraphics [width=0.593\textwidth]{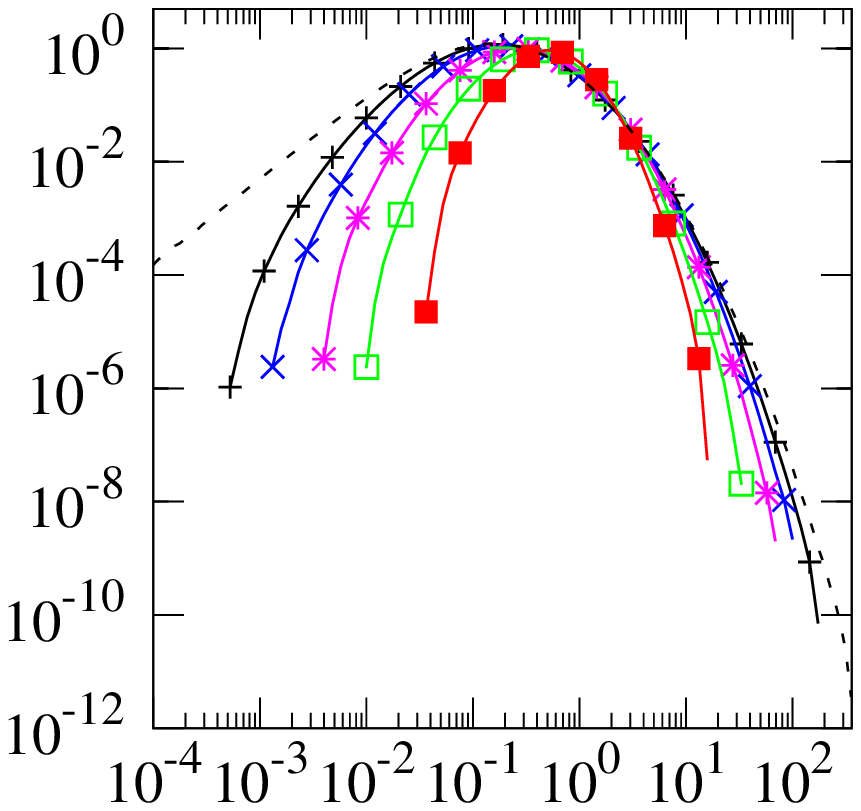}
\hspace{-5.8em}
\includegraphics [width=0.593\textwidth]{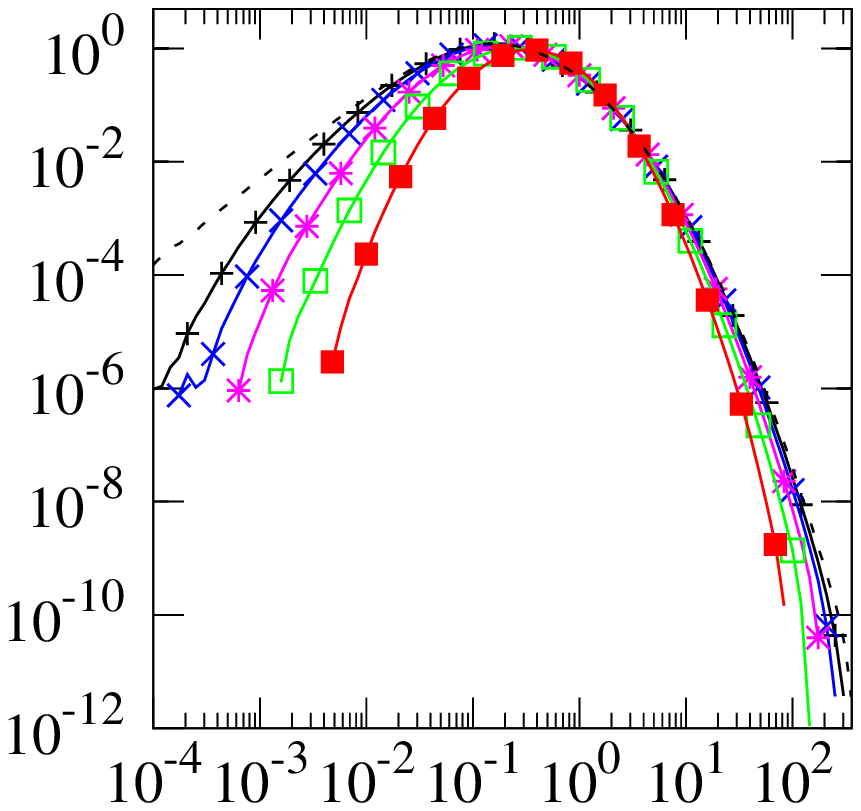}
\begin{picture}(1,1)
\put(45,2){$\epsroned/\la\epsroned\ra$}
\put(-55,2){$\epsr/\la\epsr\ra$}
\put(-115,60){\rotatebox{90}{PDF}}
\put(-115,160){\rotatebox{90}{PDF}}
\put(55,40){(d)}
\put(55,140){(b)}
\put(-45,140){(a)}
\put(-45,40){(c)}
\end{picture}
\end{minipage}
\protect\caption{(Color online) PDF of local 3D dissipation ($\epsr$)
and 1D dissipation ($\epsroned$) at $\rel \sim 240$ for
(a) $\epsr$, $\kres = 1.4$,
(b) $\epsroned$, $\kres = 1.4$,
(c) $\epsr$, $\kres = 5.7$ and
(d) $\epsroned$, $\kres = 5.7$. Curves in each frame correspond
to  
scale separations
$r/\eta \app 2$ ($+$), $4$ (\colb{$\times$}), $8$ (\colm{$\ast$}), $16$ (\colga{$\Box$}) 
and $32$ (\colr{$\blacksquare$}) respectively.
Dashed curves correspond to PDF of point-wise dissipation
($\epsilon/\meandiss$).
}
\label{pdf.fig}
\end{figure}

To compare the likelihood of extreme events of dissipation, 
we show in
Fig. \ref{pdf.fig} the PDF
for the single-point, 3D and 1D averaged dissipation 
at two different  resolutions
at the same Reynolds number.
The right-tails of the PDF which correspond to large
dissipation events get
wider with increased resolution. This effect is
more pronounced in the 1D averaged dissipation than in the
3D averaged dissipation.
In contrast, the left-tails of the PDF correspond to 
smaller dissipation
events and are less-sensitive to resolution, 
but benefit from
increased sampling in the higher resolution case. 
For a given scale size, 
the PDF of $\epsroned$ has wider tails than that of $\epsr$,
indicating that the probability of
1D dissipation taking extreme values is greater than that
of 3D dissipation. This is consistent both with the trends for the  
second order moments of dissipation
shown in
Fig.~\ref{secmom.fig} and relation \ref{dissord.eq}.
A reduced variability at a given Reynolds number can potentially
render the 3D averaged dissipation less sensitive to
finite sampling effects as compared to 1D dissipation.
\subsection{Preliminary results}
\begin{figure}[b]
\begin{minipage} [t] {0.5\textwidth}
\centering
\includegraphics[width=0.572\textwidth]{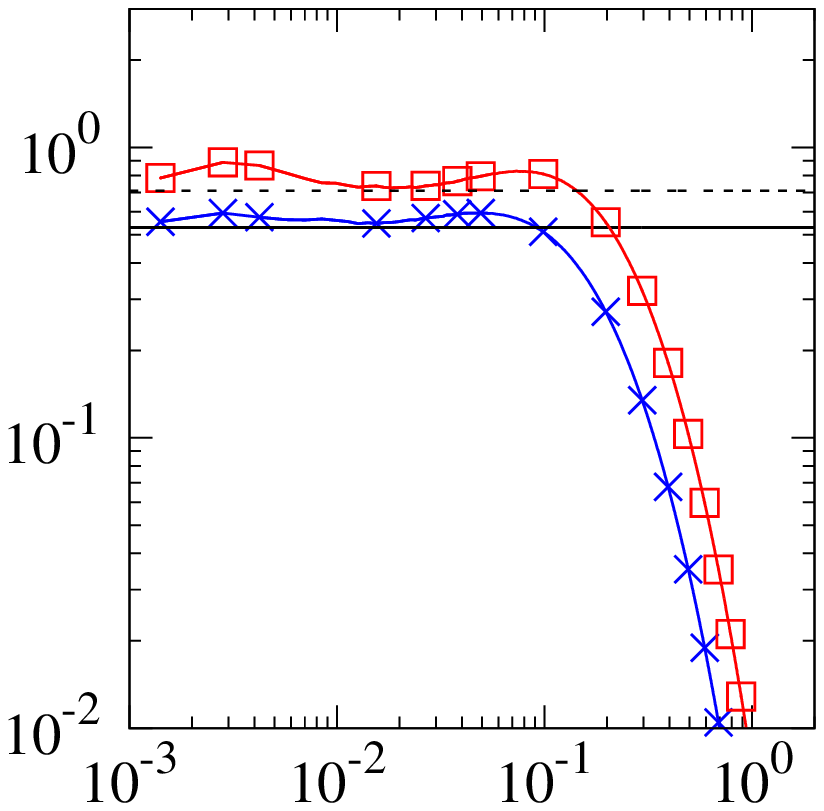}
\hspace{-4.92em}
\includegraphics[width=0.572\textwidth]{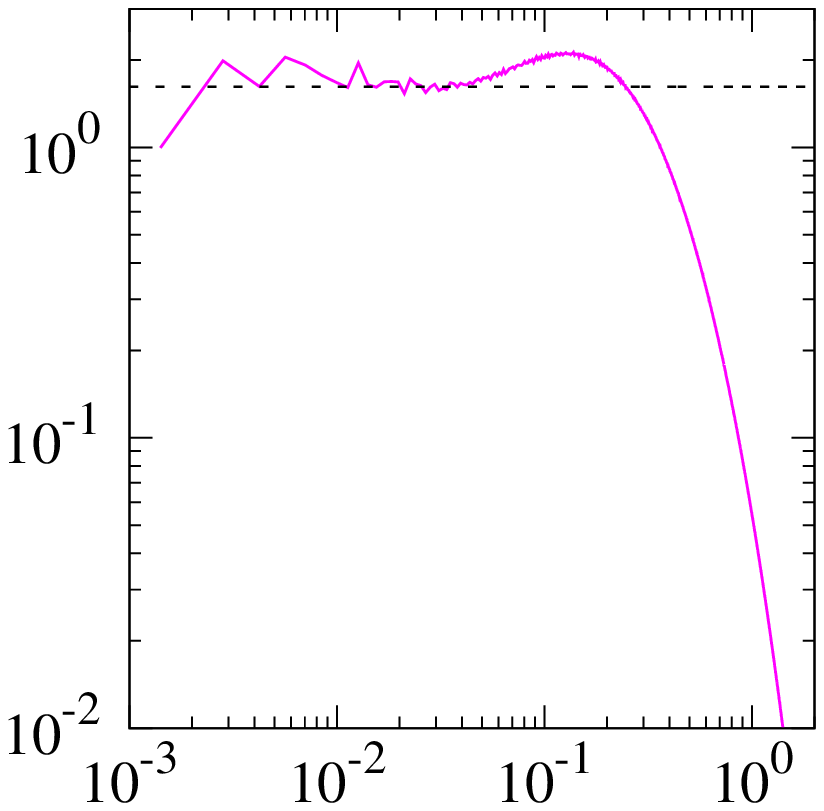}
\begin{picture}(1,1)
\put(60,1){$\keta$}
\put(-80,40){(a)}
\put(-60,1){$\koneeta$}
\put(40,40){(b)}
\put(-6,22){\rotatebox{90}{$E(k)k^{5/3}\meandiss^{-2/3}$}}
\put(-123,20){\rotatebox{90}{$E_{\alpha\alpha}(k_1)k_1^{5/3}\meandiss^{-2/3}$}}
\end{picture}
\end{minipage}
\protect\caption{(Color online) Compensated energy spectra at
$4096^3$, $\rel \sim 650$.
(a) Longitudinal (\colb{$\times$}) and transverse 
(\colr{$\Box$}) spectra as 
functions of wavenumber magnitude in $k_1$ direction,
correspond to $\alpha=1,2$ respectively. 
Solid horizontal line at $C_k = 0.53$ is the
inertial range constant for longitudinal spectrum from
experiments (Ref.~[\onlinecite{krs95}]).
Corresponding constant
for the transverse spectrum is given by the dotted line 
at $4C_k/3$.  
(b) Three-dimensional energy spectrum as a function of 
wavenumber magnitude
$k$. Dotted horizontal line at $55C_k/18$ shows the 
Kolmogorov constant. 
}
\label{ek650.fig}
\end{figure}

\begin{figure}
\begin{minipage} [t] {0.5\textwidth}
\centering
\includegraphics[width=0.604\textwidth]{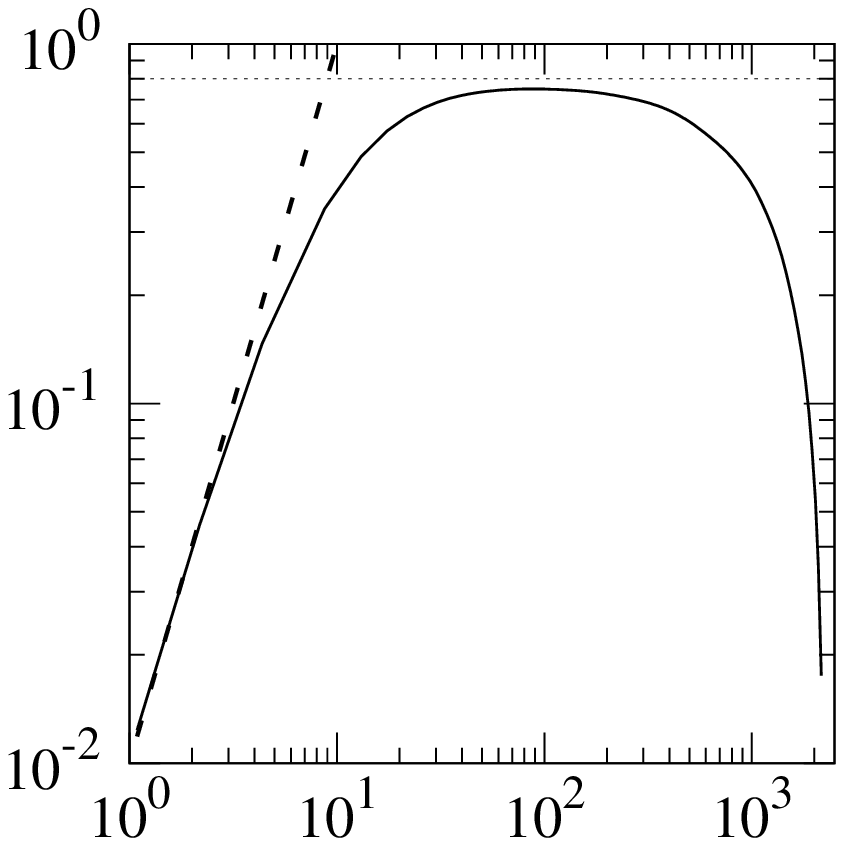}
\hspace{-6.4em}
\includegraphics[width=0.604\textwidth]{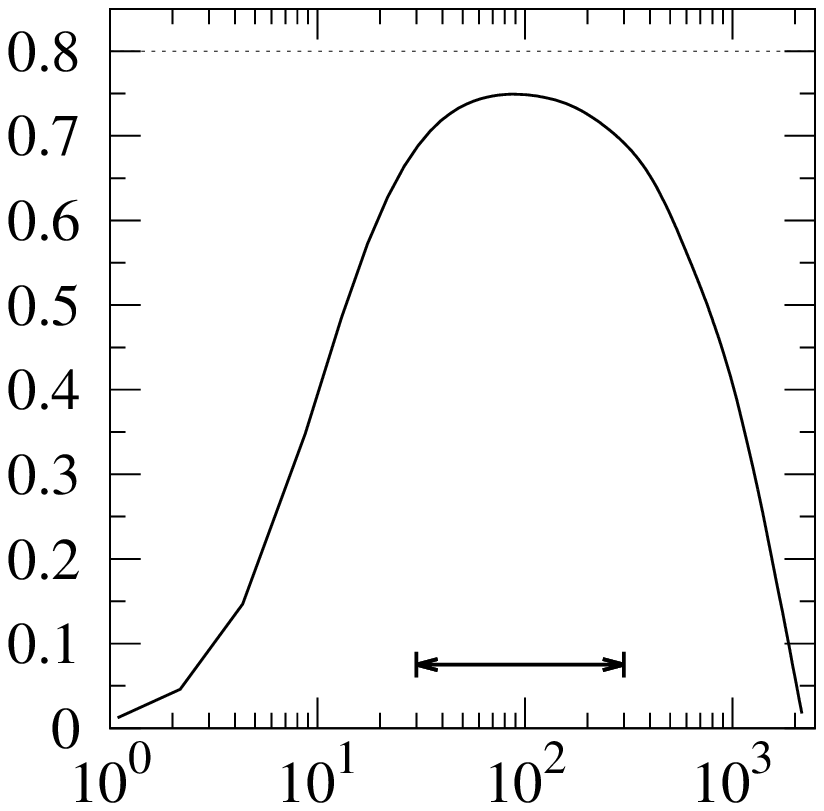}
\begin{picture}(1,1)
\put(-60,60){(a)}
\put(60,60){(b)}
\put(60,35){IR}
\put(45,2){$\normr$}
\put(-55,2){$\normr$}
\put(-118,30){\rotatebox{90}{$-\la (\Delta u(r))^3\ra/\meandiss r$}}
\end{picture}
\end{minipage}
\protect\caption{
Normalized third-order velocity structure function
at $4096^3$,
$\rel \sim 650$, 
averaged over the three Cartesian directions.
(a) Log-log scales to check small-$r$ behavior, 
dashed line corresponds to $r^2$ to check small-$r$ slope.
(b) Log-linear scales to assess inertial range extent.
Dotted line at $0.8$ for comparison with Eq.~\ref{k41.eq}.
The inertial range (IR) is taken as $30 < r/\eta < 300$.
Ratio of the integral scale to the Kolmogorov scale $L/\eta \app 1000$.
}
\label{re650dlll.fig}
\end{figure}

As a prelude to the main results, 
we provide some basic small-scale information for the 
$4096^3$ simulation.
Figure \ref{ek650.fig} shows the compensated energy spectra
at $\rel \sim 650$. The inertial range constant for the longitudinal
1D spectrum is approximately $0.53$ which is consistent with the 
conclusions
of [\onlinecite{krs95}].
Figure \ref{re650dlll.fig} shows the normalized third-order
longitudinal velocity structure function at $\rel\sim 650$.
The component averaged result (averaged over the three
Cartesian directions) shown here, is used  
to assess the extent of the inertial range.
At $\rel \sim 650$, the scale separation
is wide enough ($L/\eta \app 1000$) to
meaningfully contrast the small scales ($r \ll L$) from the 
energy-containing large scales.
The lack of convergence with the K41 plateau (refer Eq.~\ref{k41.eq})
is at least partly due to the effects of finite sampling and 
periodic boundary conditions, at the intermediate scales.
The effect of limited sampling is more pronounced at 
higher Reynolds numbers, at which
$\delur$ is known to be intermittent in the inertial range
[\onlinecite{krsh}]. For a given box length,
periodic boundary conditions influence the inertial range
statistics calculated along the Cartesian directions to a greater
extent than those calculated using other directions.
Nevertheless,
a reasonable estimate for the inertial range
for the present simulation from Fig.~\ref{re650dlll.fig}  is
$30\eta < r < 300\eta$.

\begin{figure}
\begin{minipage}[t]{0.5\textwidth}
\includegraphics [width=0.7\textwidth]{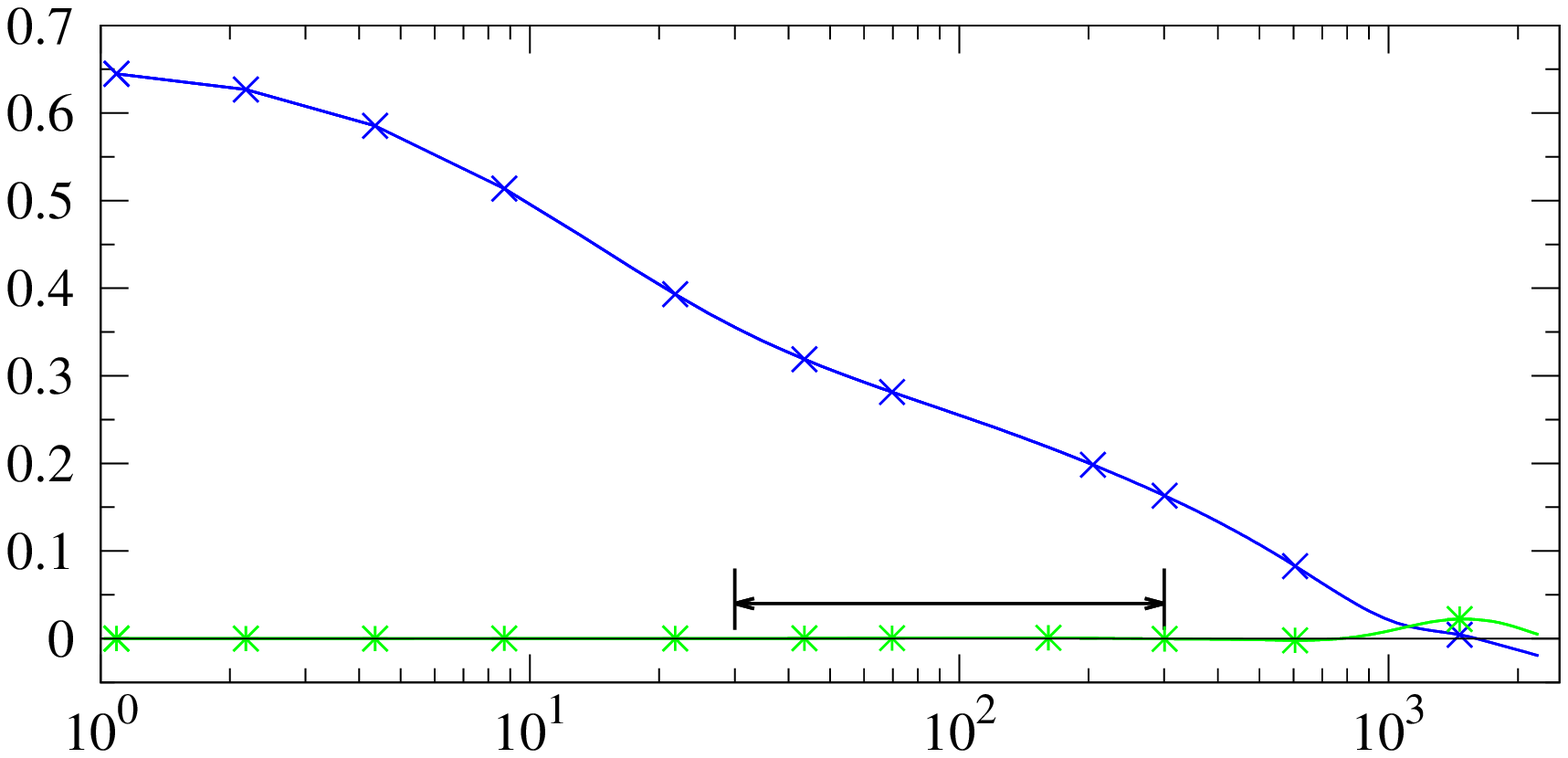}
\begin{picture}(1,1)
\put(-100,-7){$\normr$}
\put(-90,20){IR}
\put(-195,10){\rotatebox{90}{$\textrm{corr}[\beta,\repsrpow]$}}
\end{picture}
\end{minipage}
\protect\caption{(Color online)
Correlation coefficient as a function of spatial
separation between $\repsrpow$ and the quantity $\beta$ where
$\beta$ is $\delur$ (\colga{$\ast$}) and
$\modelur$ (\colb{$\times$}) for
$\rel \sim 650$, $4096^3$. 
Correlation coefficient is defined
as 
$\textrm{corr}(x,y) = 
\la (x- \la x \ra)(y-\la y \ra) \ra/\sigma_x\sigma_y$,
where $\sigma_x$ and $\sigma_y$ are the standard deviations
of $x$ and $y$. 
The inertial range
(IR) is marked for reference.
}
\label{delcorr.fig}
\end{figure}

Since we are interested in the properties of the stochastic
variable $V$, it is useful
to study the correlation between the velocity increment $\delur$ 
and the local velocity scale $\repsrpow$.
The correlation coefficient
between $\delur$ and $\repsrpow$ is 
plotted in Fig.~\ref{delcorr.fig}.
The two quantities are insignificantly correlated at all
spatial separations as a consequence of isotropy. 
The minor deviations at the largest scales
may be due to anisotropic
effects of forcing and 
finite domain size considerations.
Considering $\modelur$ instead of $\delur$ results in stronger
correlation (Fig.~\ref{delcorr.fig}) at the small scales.
Figure \ref{delcorr.fig} shows that the correlation
coefficient between $\modelur$ and $\repsrpow$ varies
between $0.15$ and
$0.35$ in the inertial range.
These results are consistent with those reported in
Ref.~[\onlinecite{krsh}] 
and serve as useful checks on the validity
of the data. 
\subsection{K62: the second hypothesis}
The second K62 hypothesis is that at sufficiently large
local Reynolds number, 
the PDF of $V(r)$ for $\eta \ll r \ll L$ becomes 
independent of $\rer$ and hence is
universal.
In our simulations, the viscosity is a constant. Hence it
suffices to check for the dependence of $V$ on the
local velocity scale $\repsrpow$ and the separation length $r$.

\begin{figure}[!]
\begin{minipage} [t] {0.5\textwidth}
\includegraphics [width=0.4\textwidth]{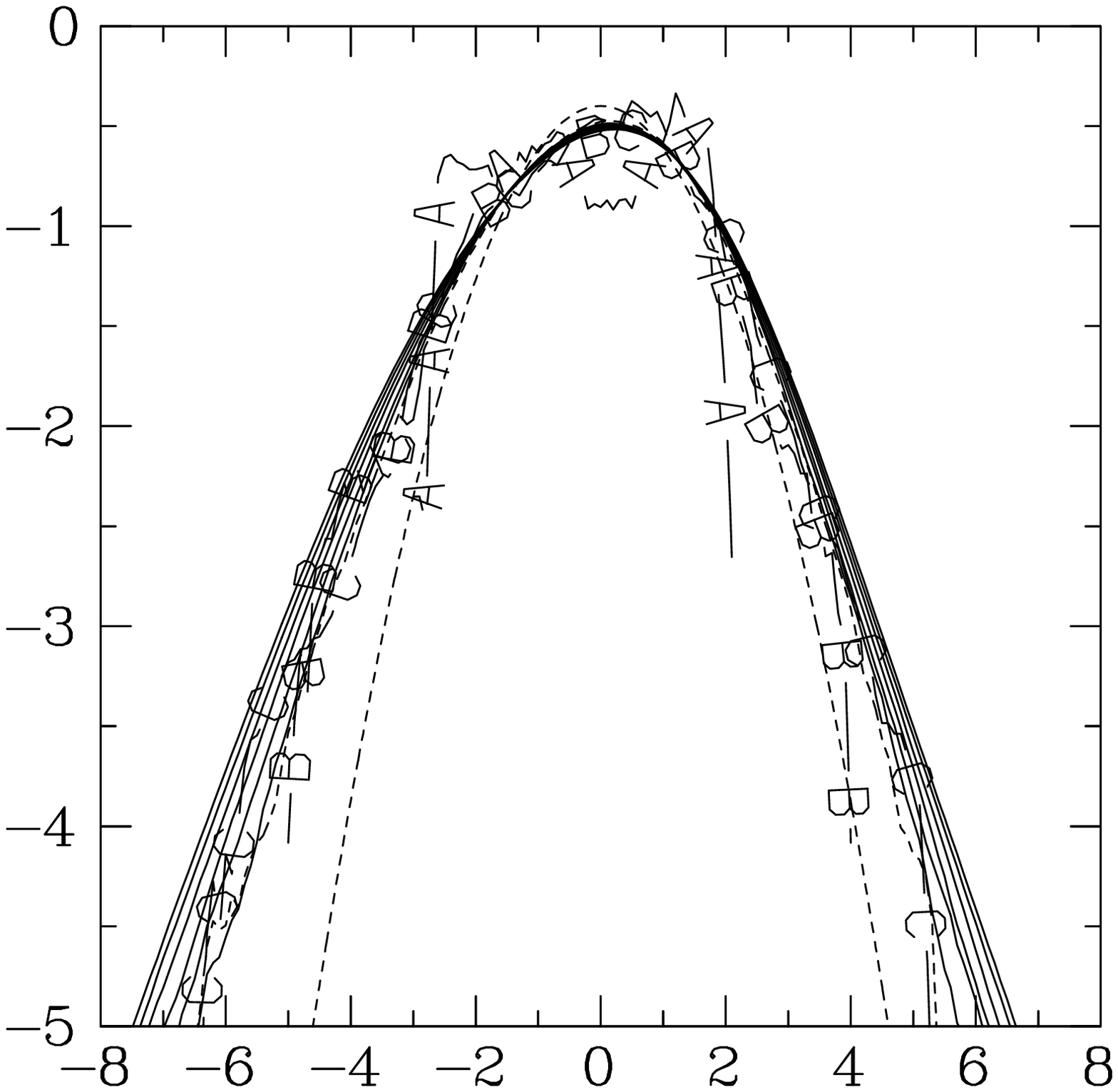}
\includegraphics [width=0.4\textwidth]{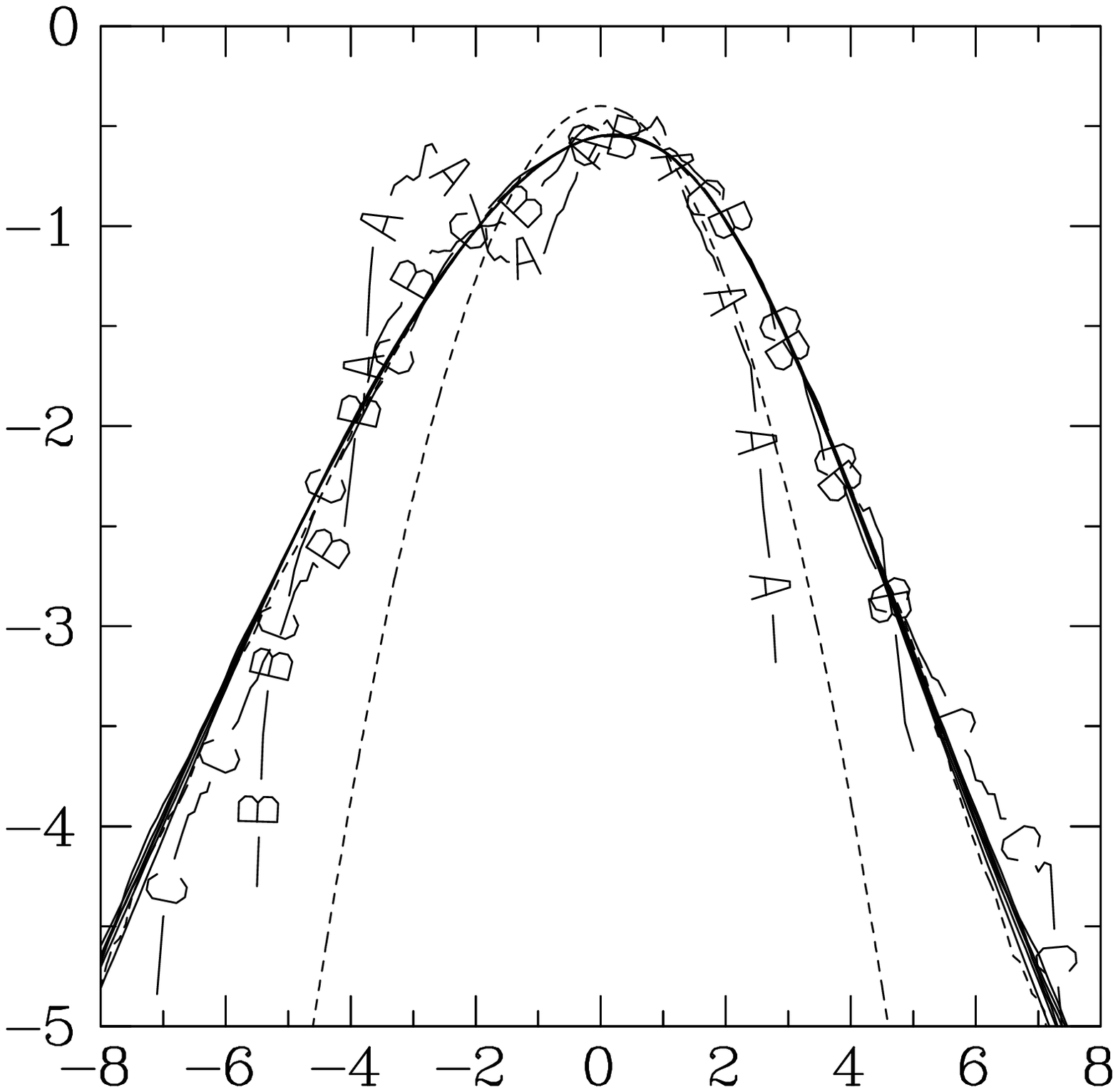}\\
\includegraphics [width=0.4\textwidth]{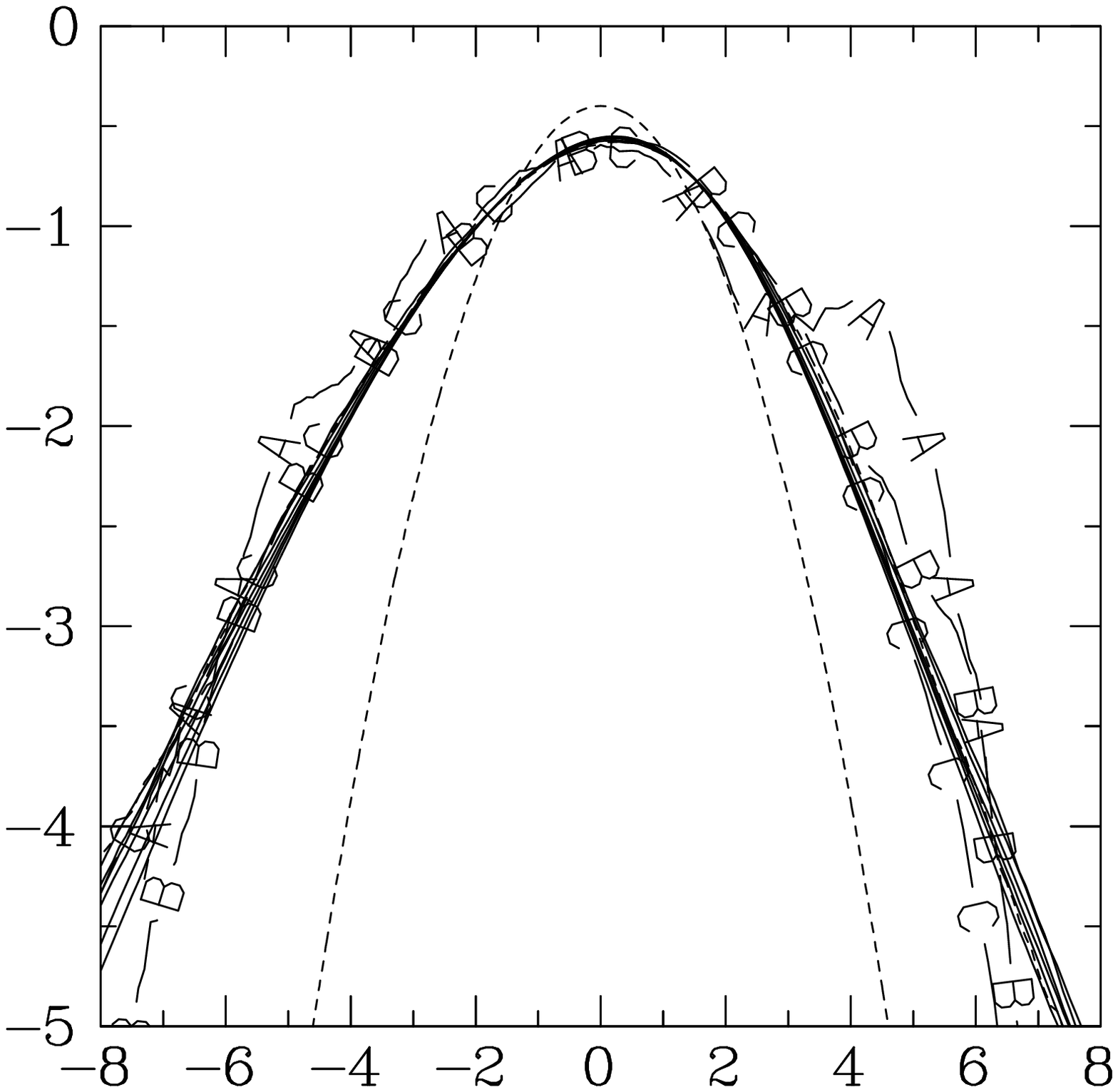}
\includegraphics [width=0.4\textwidth]{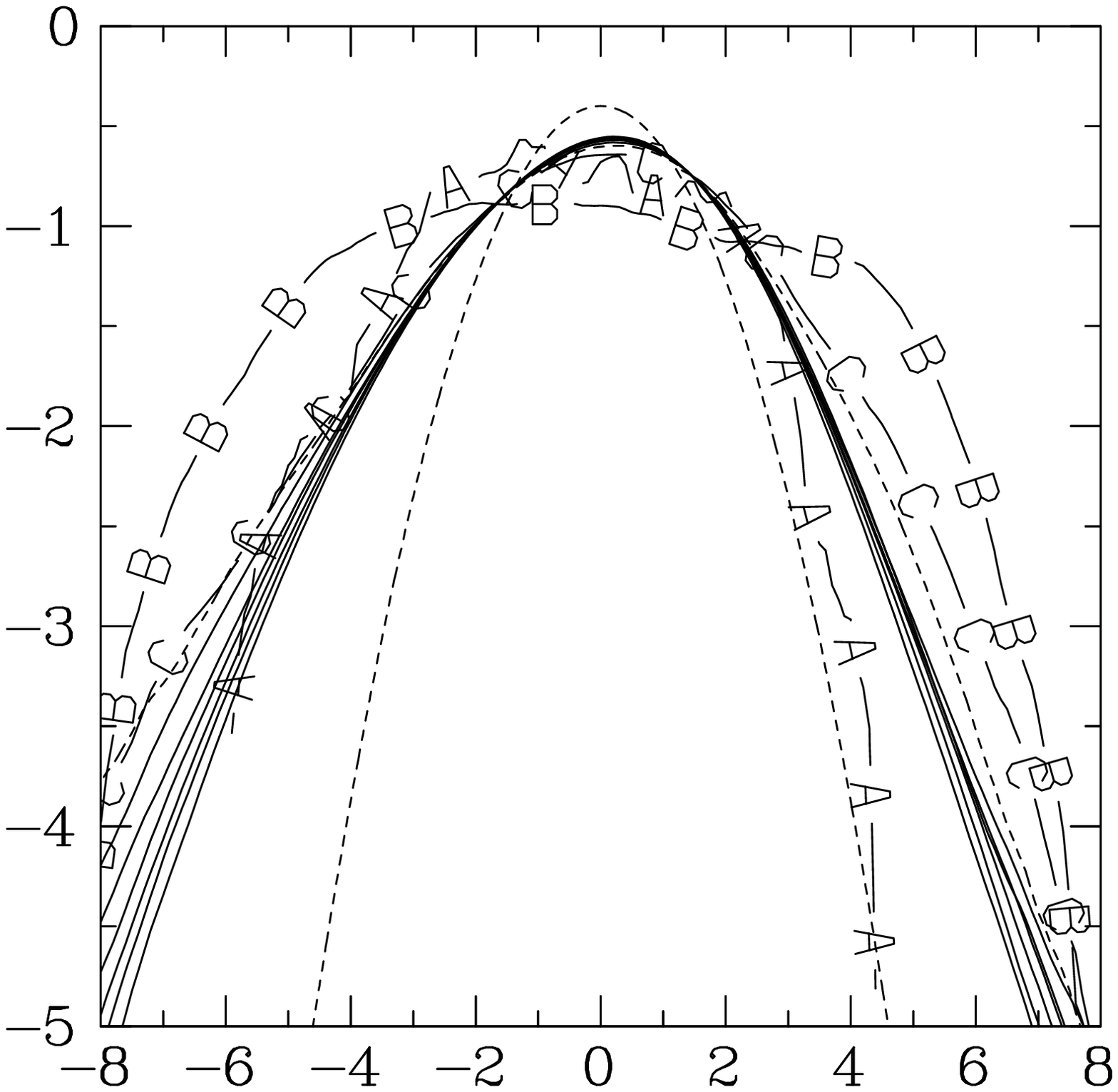}
\vspace{5mm}
\begin{picture}(1,1)
\put(-175,-15){$V(r)$}
\put(-70,-15){$V(r)$}
\put(-230,105){\rotatebox{90}{$\log_{10} P[V(r)|\repsrpow] $}}
\put(-230,5){\rotatebox{90}{$\log_{10} P[V(r)|\repsrpow] $}}
\put(-160,135){(a)}
\put(-170,125){$\normr=35$}
\put(-60,135){(b)}
\put(-65,125){$\normr=70$}
\put(-160,25){(c)}
\put(-175,15){$\normr=139$}
\put(-60,25){(d)}
\put(-69,15){$\normr=279$}
\end{picture}
\end{minipage}
\protect\caption{
Conditional PDF of $V(r)$
for a given spatial separation.
The separation distance $\normr$, the number of curves, and the
minimum and maximum $\rer$ in each frame are
as follows:
(a) $35$, $10$, $21$ and $71$.
(b) $70$, $10$, $71$ and $247$.
(c) $139$,$10$, $260$ and $595$.
(d) $279$, $10$, $823$ and $1886$.
Curves $A$, $B$ and $C$ in each frame correspond
to the three lowest values of $\rer$ (in ascending order).
In (b) and (c), where inertial effects dominate
(Fig.~\ref{re650dlll.fig}), the three
 labelled curves corresponding to the three lowest $\rer$
 are distinct from the others (at higher $\rer$) which collapse. 
In frames (a) and (d) where non-inertial effects are more
significant, the collapse of the unmarked curves
is less pronounced.
Dashed curve is the Gaussian distribution with zero mean and unity variance.
}
\label{viner.fig}
\end{figure}

Figure \ref{viner.fig} shows the
PDF of $V$ conditioned on $\repsrpow$ for
four separation distances.
In each frame of Fig.~\ref{viner.fig}, the PDFs marked as
$A$, $B$ and $C$ correspond to the lowest $\rer$ and are distinct
from the unmarked PDFs which correspond to higher $\rer$, which
coalesce, with the shape being preserved in going from one separation
distance to another. The collapse of the PDFs of $V$ at higher $\rer$, 
indicates an approximate
independence of $V$ from $\epsr$ and $r$ in the inertial range.
The universality of $V$ as evidenced by the collapse of the
PDFs at high enough $\rer$ is more pronounced at the intermediate
separations
(frames (b) and (c) in Fig.~\ref{viner.fig}),
where non-inertial
effects such as viscosity and large-scale forcing are less important
(Fig.~\ref{re650dlll.fig}).

\begin{figure}[b]
\begin{minipage} [t] {0.5\textwidth}
\includegraphics [width=0.7\textwidth]{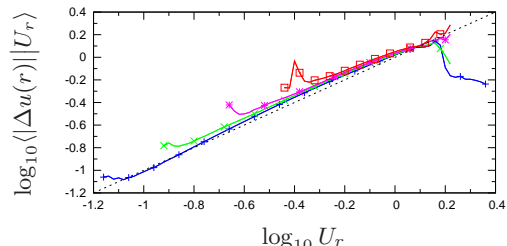}
\begin{picture}(1,1)
\put(-100,-10){$\log_{10} U_r$}
\put(-195,8){\rotatebox{90}{$\log_{10} \la |\Delta u(r)| \big|U_r \ra$}}
\end{picture}
\end{minipage}
\protect\caption{(Color online)
Logarithm of the mean of $|\Delta u(r)|$ conditioned on
the local velocity scale $U_r = \repsrpow$ as a 
function of the logarithm of $U_r$.
Symbols (\colb{$+$}), (\colga{$\times$}), (\colm{$\ast$}), 
(\colr{$\Box$})
correspond to inertial range separations 
$\normr = 35,70,139$ and $279$, respectively.
Dashed line has a slope of $1$.
}
\label{condexpiner.fig}
\end{figure}

In order to further test the dependence of $V$ on $\repsrpow$, 
we consider the mean value $\modelur$ conditioned on 
$\repsrpow$. It follows from Eq.~\ref{k621.eq} that
\beq
\label{cond.eq}
\la \modelur \bigd \repsrpow \ra =
\repsrpow \la \modv \bigd \repsrpow \ra
\eeq
If $V$ were independent of $\repsrpow$ in the inertial range, it is clear that
the left-hand side of Eq.~\ref{cond.eq} would be a linear function of
$\repsrpow$ for all values of $r$ in the inertial range. In such a 
scenario, Eq.~\ref{cond.eq}
becomes
\beq 
\label{condlast.eq}
\la \modelur \bigd \repsrpow \ra
= \repsrpow \la \modv \ra \;,\quad \eta \ll r \ll L\;.
\eeq
Figure \ref{condexpiner.fig} shows that this indeed is the case, 
except possibly at the tails where
the sampling uncertainty can be large. This confirms that
$\modv$ as well as $V$ are independent of 
$\repsrpow$ in the inertial range.

\begin{figure}[!]
\begin{minipage}[t]{0.5\textwidth}
\includegraphics [width=0.7\textwidth]{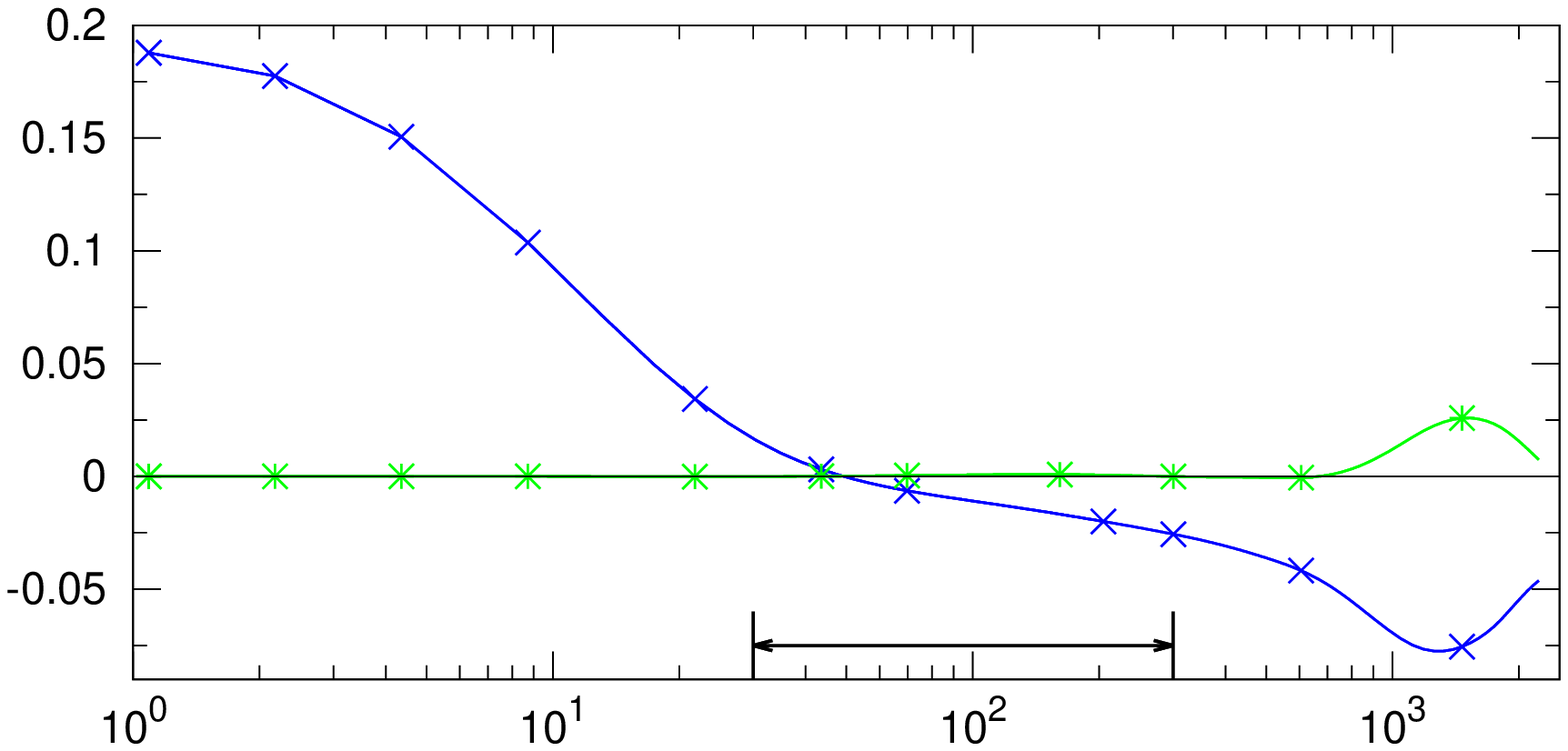}
\begin{picture}(1,1)
\put(-100,-7){$\normr$}
\put(-92,15){IR}
\put(-195,10){\rotatebox{90}{$\textrm{corr}[\gamma,\repsrpow]$}}
\end{picture}
\end{minipage}
\protect\caption{(Color online)
Correlation coefficient
 as a function of spatial
separation between $\repsrpow$ and the quantity $\gamma$ where
$\gamma$ is either $V(r)$ (\colga{$\ast$}) or 
$|V(r)|$ (\colb{$\times$}), at
$\rel \sim 650$, $4096^3$. 
The inertial range (IR) is marked for
reference.
}
\label{vcorr.fig}
\end{figure}

Since $V$  is approximately independent of $\repsrpow$ in the 
inertial range,
it follows that the correlation between $\modv$ and $\repsrpow$ in this scale
range should be zero.
Figure \ref{vcorr.fig} shows that the correlation coefficient between
$\modv$ and $\repsrpow$ is indeed close to zero in the 
inertial range. 
The correlation between $V$ and $\repsrpow$ in Fig. ~\ref{vcorr.fig} 
is trivially
zero
due to homogeneity.
At the largest scales, the non-zero correlation between
$V$ and $\repsrpow$ may 
be due to
the finite size of the domain.

\begin{figure}[!]
\begin{minipage}[t]{0.5\textwidth}
\includegraphics [width=0.604\textwidth]{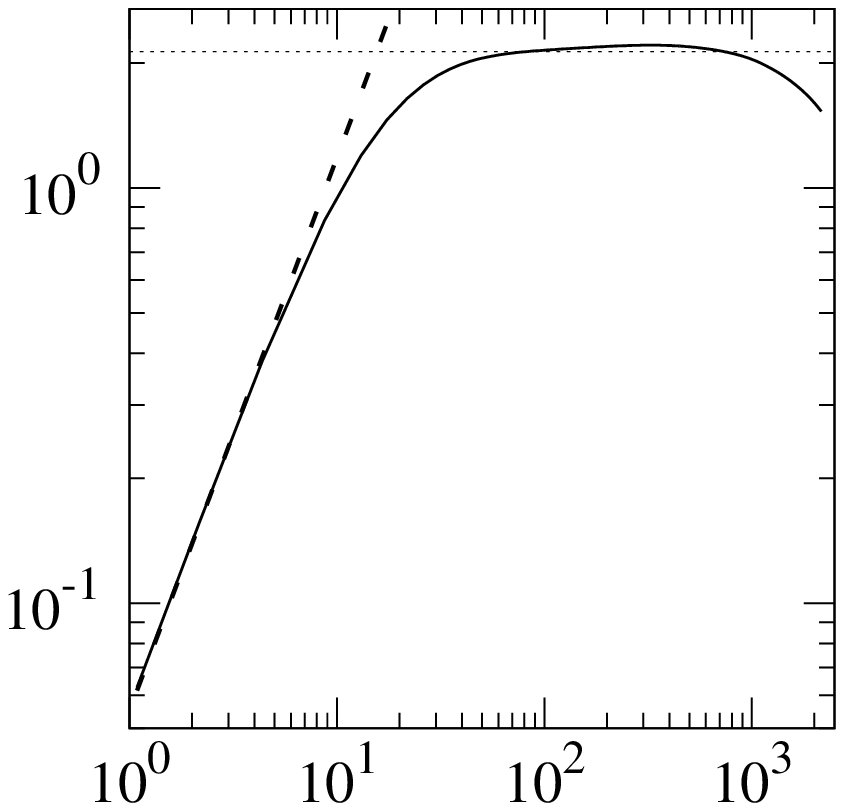}
\hspace{-6.4em}
\includegraphics [width=0.604\textwidth]{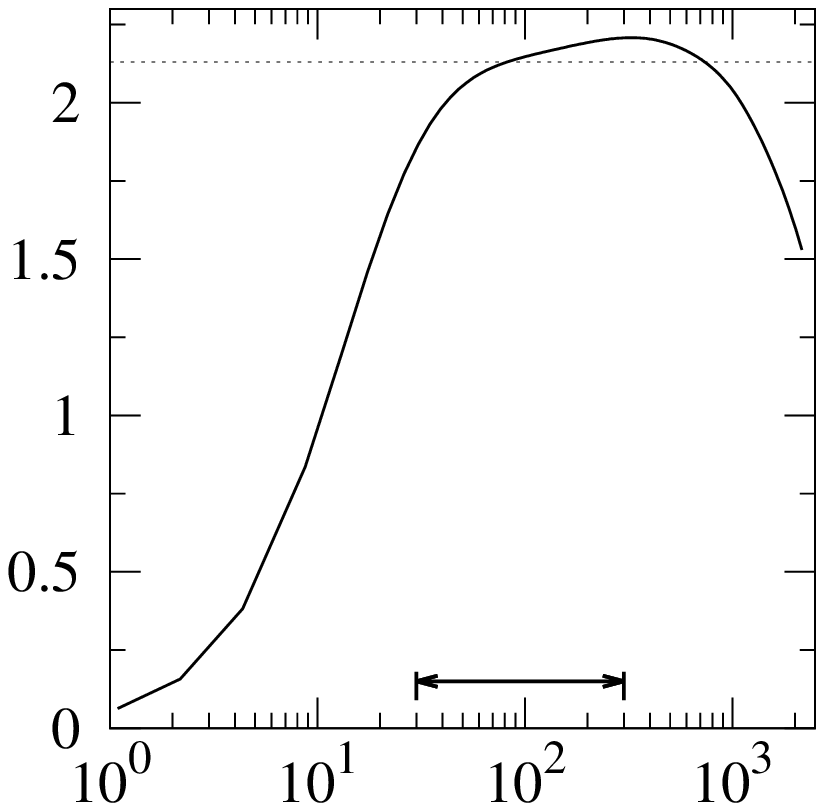}
\begin{picture}(1,1)
\put(-60,60){(a)}
\put(60,60){(b)}
\put(60,35){IR}
\put(45,2){$\normr$}
\put(-55,2){$\normr$}
\put(-118,45){\rotatebox{90}{$\la (V(r))^2 \ra$}}
\end{picture}
\end{minipage}
\protect\caption{
Second moment of $V$
at $4096^3$, $\rel \sim 650$.
(a) Log-log scales to verify small-$r$ behavior, 
dashed line corresponds to $r^{4/3}$
to check the small-$r$ slope. 
(b) Log-linear scales to check inertial range behavior; 
Dotted line at $2.13$ for comparison with corresponding
$K41$ result.
The inertial range (IR) extent from Fig.~\ref{re650dlll.fig}
is shown for reference.
}
\label{re650v2.fig}
\end{figure}

The above results show that the stochastic variable $V$ in the inertial
range  is
independent of $r$ and $\epsr$, and that it is
approximately universal. It then
follows from Eq.~\ref{k621.eq} that the $m$-th order 
structure function in the inertial range is given as
\beq
\label{mthord.eq}
\la [\delur]^m \ra = \la V^m \ra \la (r\epsr)^{(m/3)} \ra \;.
\eeq
In particular, $\la V(r) \ra = 0$ in the inertial range.
The second and third moments of $V$ in the inertial range are related
to the corresponding longitudinal velocity structure functions by
\beqa
\label{v2mom.eq}
\la(\delur)^2\ra &=& \la V^2 \ra \la\epsr^{2/3}\ra r^{2/3} \;, \\
\vspace{2mm}
\label{v3mom.eq}
\la(\delur)^3\ra &=& \la V^3 \ra \la\epsr\ra r \;.
\eeqa
Figure \ref{re650v2.fig} shows the second moment of $V$
 as a function
of spatial separation. In the inertial range, $\la V^2\ra \approx 2.13$
(Fig.~\ref{re650v2.fig} (b)),
which is comparable to the accepted estimates of the 
Kolmogorov constant in second-order structure functions
(Refs.~[\onlinecite{pkzhou,MY.II}]). 
In the small-$r$ limit, a Taylor
expansion shows that $\la V^2 \ra$ varies as $r^{4/3}$, which
is confirmed in Fig.~\ref{re650v2.fig} (a). 

\begin{figure}[!]
\begin{minipage}[t]{0.5\textwidth}
\centering
\includegraphics [width=0.604\textwidth]{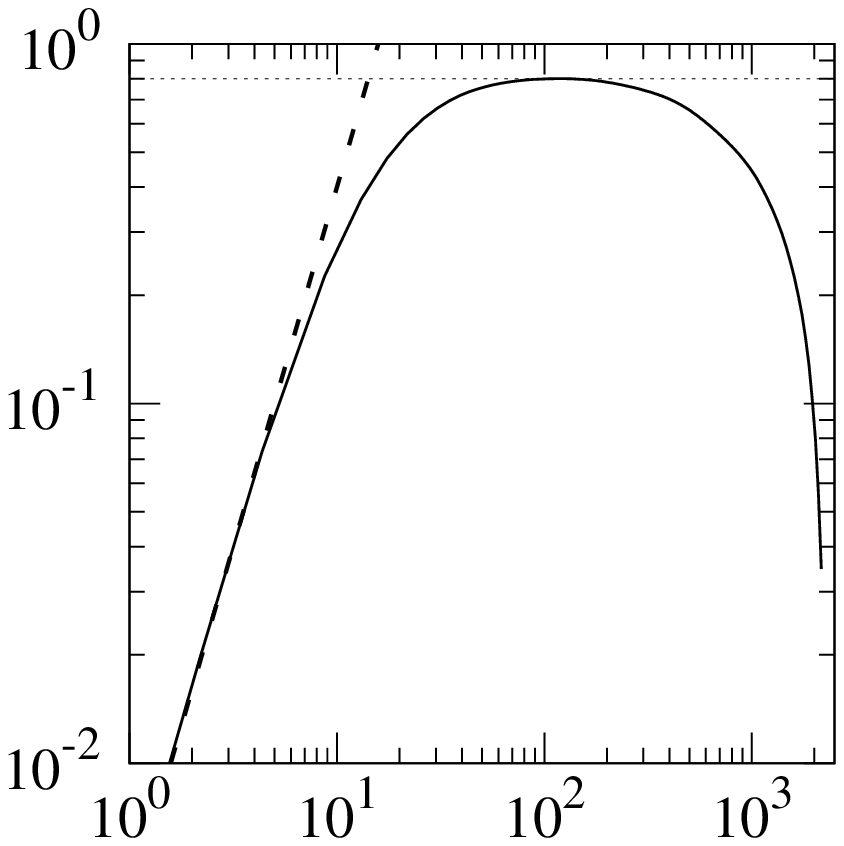}
\hspace{-6.4em}
\includegraphics [width=0.604\textwidth]{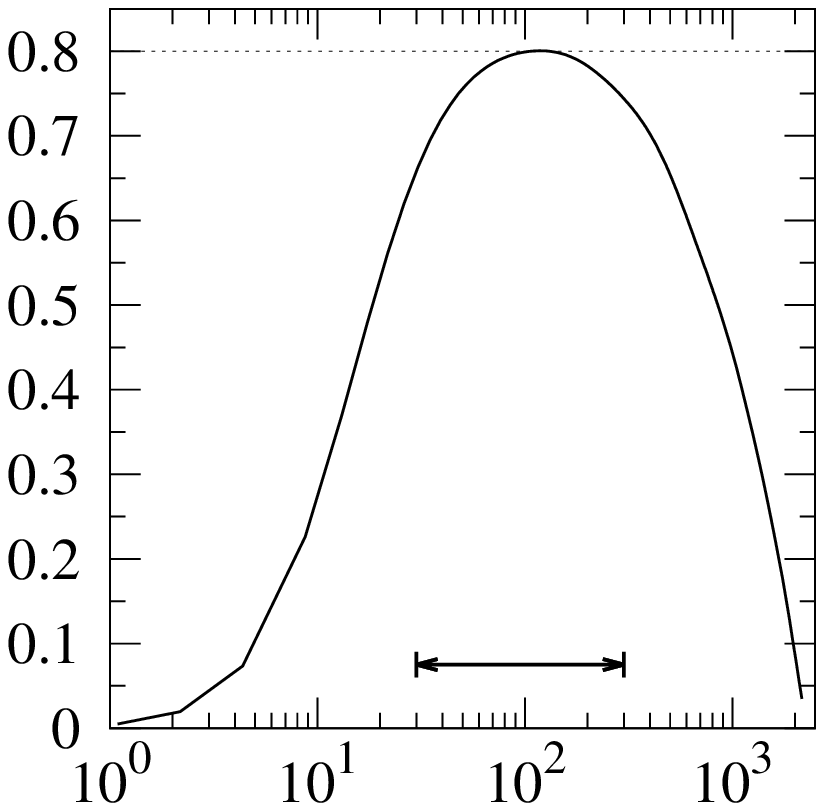}
\begin{picture}(1,1)
\put(-60,60){(a)}
\put(60,60){(b)}
\put(60,35){IR}
\put(45,2){$\normr$}
\put(-55,2){$\normr$}
\put(-118,45){\rotatebox{90}{$-\la (V(r))^3 \ra$}}
\end{picture}
\end{minipage}
\protect\caption{
Third moment of $V$
at $4096^3$, $\rel \sim 650$ (Eq.~\ref{v3mom.eq}).
(a) Log-log scales to check small-$r$ behavior, 
dashed line corresponds to $r^2$ to
check small-$r$ slope. 
(b) Log-linear scales to check inertial range behavior. 
Dotted line at $0.8$ shows $K41$ plateau.
The inertial range (IR) extent from Fig.~\ref{re650dlll.fig}
is shown for reference.
}
\label{re650v3.fig}
\end{figure}

Comparing Eqs.~\ref{v3mom.eq} and \ref{k41.eq}, we see that
$\la V^3 \ra$ should be negative in the inertial range with a
magnitude of $4/5$.
Figure \ref{re650v3.fig} shows the third moment of $V$
(Eq.~\ref{v3mom.eq}) at $\rel \sim 650$ as a function of
spatial separation.
In the inertial range,
$\la V^3 \ra = -0.8$ (Fig.~\ref{re650v3.fig} (b)), 
which is consistent with
the exact $K41$ result (Eq.~\ref{k41.eq}). 
The quadratic form of $\la V^3\ra$ at $r \app \eta$ is confirmed
in Fig.~\ref{re650v3.fig} (a) which shows a $r^2$ behavior at the
smallest scales.  

The stochastic variable $V$ is universal in the 
inertial range, in particular it is independent of $r$ and 
$\epsr$ in this scale range.
Hence it follows from
Eq.~\ref{k621.eq} that
$\var(V^3) \la r\epsr \ra^2 < \var (\delur^3)$,  
where $\var(\cdot)$ denotes the variance of a random variable.
Using homogeneity and
dissipative anomaly 
[\onlinecite{dissana}], 
we get
\beq
\var(V^3) < \var \Big ( \frac{\delur^3}{r\meandiss} \Big ) \;\;,
\quad \textrm{for} \;\;\; \eta \ll r \ll L,
\eeq
Even though $V$ is the ratio of two intermittent quantities, it
is well defined at least in the inertial range and that
$V$ is a better estimator of the K41 constant
than $\delur$. This is consistent with Figs.~\ref{re650dlll.fig}
and \ref{re650v3.fig} which show that $\la V^3 \ra$ converges
faster than $\la (\Delta u(r))^3 \ra$ to the $4/5$th plateau. 

\subsection{K62: the first hypothesis}
The statement of the first K62 postulate is that for 
$r \ll L$, the PDF of $V$ depends only on $\rer$. Since the 
viscosity is constant in our simulation, it is sufficient
to check the dependence of $V$ on
$r\repsrpow$.

\begin{figure}[!]
\begin{minipage}[t]{0.5\textwidth}
\includegraphics [width=0.604\textwidth]{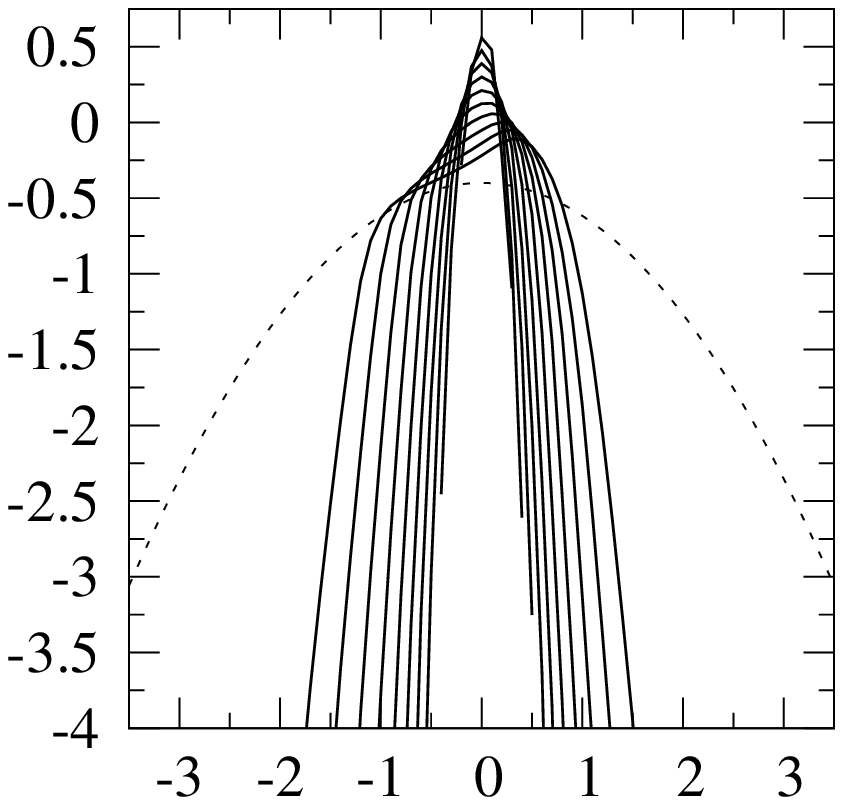}
\hspace{-6.4em}
\includegraphics [width=0.604\textwidth]{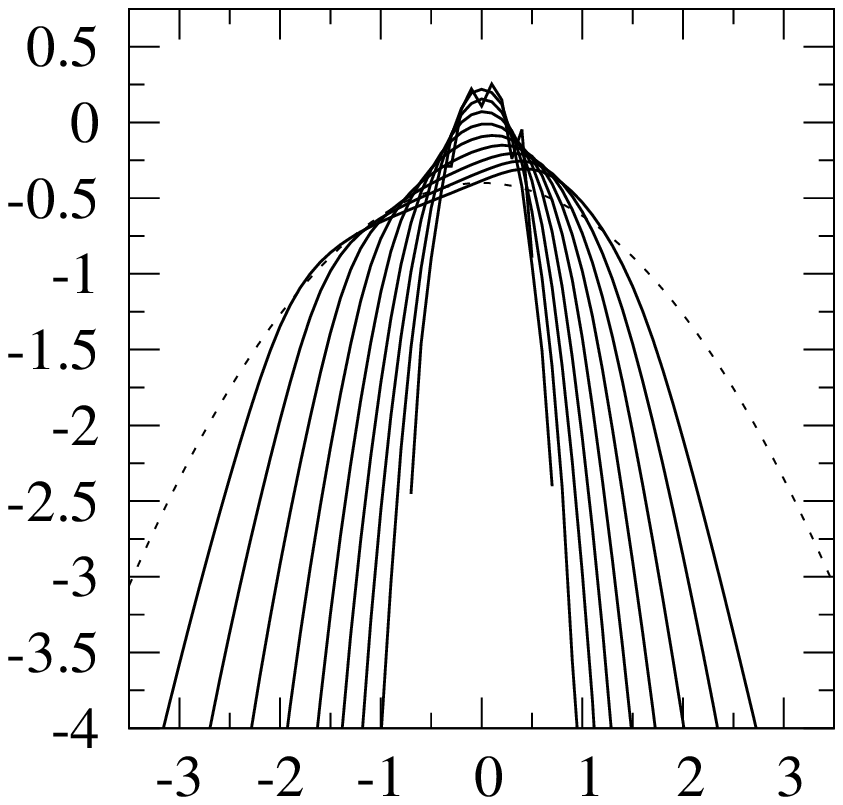}
\begin{picture}(1,1)
\put(-80,100){(a)}
\put(20,100){(b)}
\put(-85,90){$\normr=2$}
\put(15,90){$\normr=4$}
\put(51,25){\vector(-1,2){25}}
\put(55,25){\vector(1,2){25}}
\put(-48,25){\vector(-1,2){20}}
\put(-40,25){\vector(1,2){20}}
\put(45,2){$V(r)$}
\put(-55,2){$V(r)$}
\put(-118,20){\rotatebox{90}{$\log_{10} P[V(r)|\repsrpow] $}}
\end{picture}
\end{minipage}
\protect\caption{
Conditional PDF of $V(r)$ in the small-scale range for a given 
spatial separation.
The separation distance $r$ (fixed for each frame),
and the minimum and maximum 
values of $\rer$ are as follows:
(a) $\normr = 2$, $\rer$
ranging from $0.2$ to $4.4$.
(b) $\normr = 4$, 
$\rer$ ranging from $0.5$ to
$12.9$.
In each frame, there are $10$ curves, each corresponding to 
a different $\rer$, increasing in the direction shown.
Dashed curve is the standard Gaussian distribution.
}
\label{vsmallrpdf.fig}
\end{figure}

Figure \ref{vsmallrpdf.fig} presents the PDF of $V$ for
two different spatial separations in the small scale regime.
Different curves in each panel corresponds to
different values of $\repsrpow$. The uncertainty in the data
is appreciably higher than in the corresponding inertial range
PDFs (Fig.~\ref{viner.fig}), because the averaging
intervals are smaller.
Even so, the conclusion from Fig.~\ref{vsmallrpdf.fig} is that the
PDF of $V$ depends on $\repsrpow$ and $r$. 

\begin{figure}[!]
\begin{minipage}[t]{0.5\textwidth}
\includegraphics [width=0.7\textwidth]{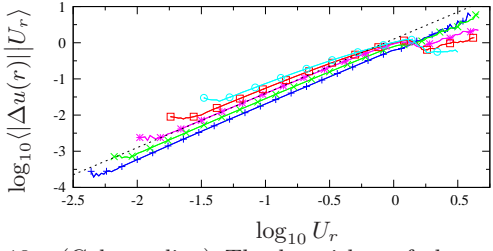}
\begin{picture}(1,1)
\put(-100,-10){$\log_{10} U_r$}
\put(-195,7){\rotatebox{90}{$\log_{10}\la\modelur \bigd U_r\ra $}}
\end{picture}
\end{minipage}
\protect\caption{(Color online) 
The logarithm of the expectation $\modelur$
conditioned on the velocity scale 
$U_r =\repsrpow$,
as a function of the logarithm of $U_r$
for different spatial separations in the small $r$ range.
Symbols (\colb{$+$}), 
(\colga{$\times$}), (\colm{$\ast$}),
(\colr{$\Box$}) and (\colc{$\bigcirc$}) correspond to
spatial separations $\normr=1,2,4,8$ and $16$, respectively. 
Dashed line has slope of $1.5$.
}
\label{smallscexp1.fig}
\end{figure}

As a further test, Fig.~\ref{smallscexp1.fig} shows the 
logarithm of the 
conditional mean of  $\modelur$, conditioned on the local
velocity as a function of the logarithm of the
local velocity in the 
small $r$ range. Evidently the curves for different
spatial separations at the smallest scales do not collapse,
confirming the dependence of $V$ on $\repsrpow$ and $r$.

\begin{figure}[!]
\begin{minipage}[t]{0.5\textwidth}
\includegraphics [width=0.604\textwidth]{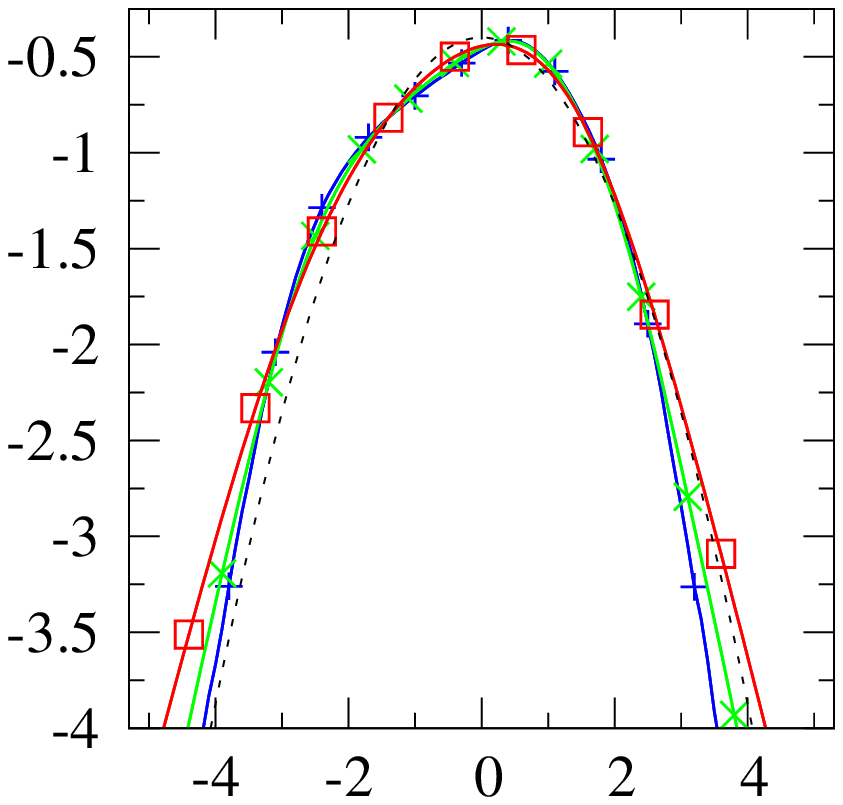}
\hspace{-6.4em}
\includegraphics [width=0.604\textwidth]{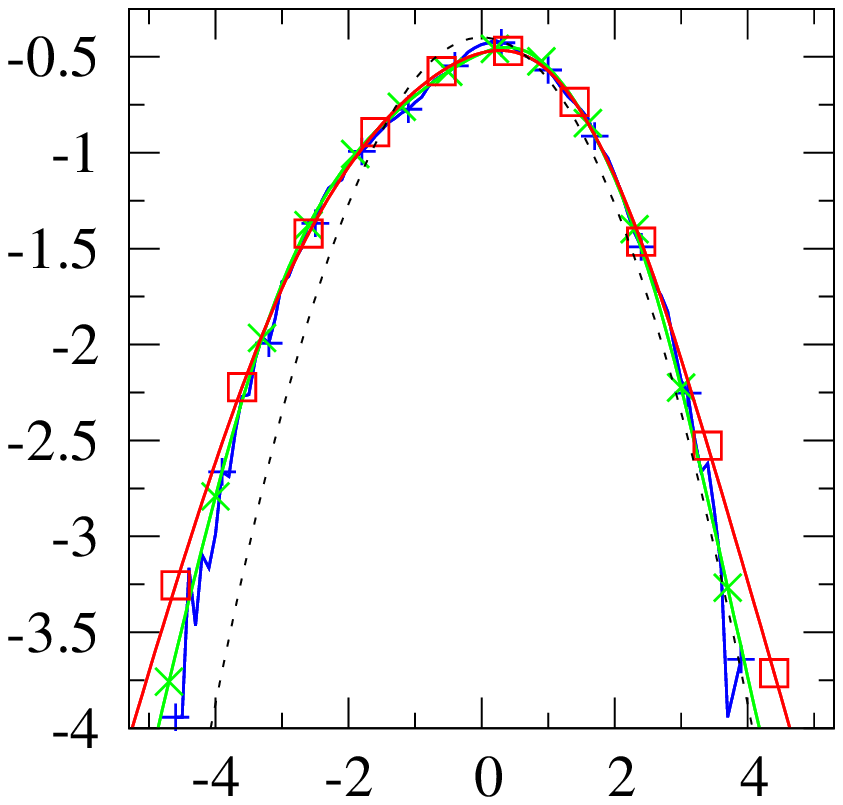}
\begin{picture}(1,1)
\put(-50,60){(a)}
\put(50,60){(b)}
\put(-60,50){$\rer \app 27$}
\put(40,50){$\rer \app 39$}
\put(45,2){$V(r)$}
\put(-55,2){$V(r)$}
\put(-118,20){\rotatebox{90}{$\log_{10} P[V(r)|\repsrpow] $}}
\end{picture}
\end{minipage}
\protect\caption{(Color online)
Conditional PDF of $V(r)$ for a given 
local Reynolds number $\rer$, for different spatial 
separations in the small-scale range.
(a) $\rer \app 27$ and (b) $\rer \app 39$.
Symbols (\colb{$+$}), 
(\colga{$\times$}), (\colr{$\Box$}) correspond to
spatial separations $\normr=4,8$ and $17$, respectively. 
The exact values of $\rer$ are within $12\%$ of each other 
in each panel. 
Dashed curve is the standard Gaussian distribution. 
}
\label{smallruniv.fig}
\end{figure}

In order to directly test the first hypothesis,
Fig.~\ref{smallruniv.fig} reports the
conditional PDF of $V$ for a fixed local Reynolds number ($\rer$). 
The PDFs in each frame correspond to different spatial separations ($r$)
and local velocities ($\repsrpow$) such that the local Reynolds number $\rer$ is 
approximately the same.
Since $\epsr$ is a random variable, exact correspondence in the
values of $\rer$ is difficult, hence $\rer$ that are within
$12\%$ of each other are considered as approximately equal in this
analysis.
The PDFs collapse onto each other with some differences at the tails
which can be attributed, at least in part
to sampling uncertainties. 
From Fig.~\ref{smallruniv.fig},
it can be concluded that the PDF of $V(r)$
only depends on the local Reynolds number $\rer$ for $r \ll L$.

\begin{figure}[!]
\begin{minipage}[t]{0.5\textwidth}
\includegraphics [width=0.7\textwidth]{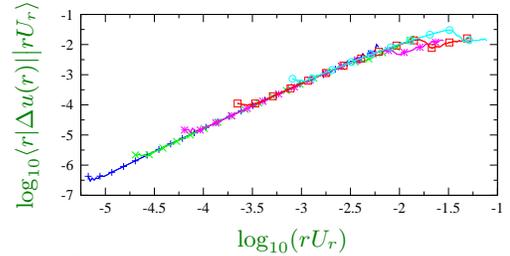}
\begin{picture}(1,1)
\put(-110,-10){\colg{$\log_{10}(r U_r)$}}
\put(-195,4){\rotatebox{90}{\colg{$\log_{10}\la r\modelur \bigd r U_r\ra $}}}
\end{picture}
\end{minipage}
\protect\caption{(Color online)
The logarithm of the expectation of 
 $r\modelur$
conditioned on
$r U_r$  ($U_r = \repsrpow$),
as a function
of the logarithm of $r U_r$
for different spatial separations in the small $r$ range.
Symbols (\colb{$+$}), 
(\colga{$\times$}), (\colm{$\ast$}),
(\colr{$\Box$}) and (\colc{$\bigcirc$}) correspond to
spatial separations $\normr=1,2,4,8$ and $16$, respectively. 
}
\label{smallscexp2.fig}
\end{figure}

In order to further test the dependence of $V$ on the local
Reynolds number in the small-scale range,
consider the mean of $r\modelur$, conditioned on
$r\repsrpow$,
\beq
\label{cond3.eq}
\la r\modelur \bigd r\repsrpow \ra =
 r\repsrpow \la \modv \bigd r\repsrpow \ra.
\eeq
If $V$ were only a function of $\rer$, or $r\repsrpow$
for $r \ll L$, then
the
above equation becomes
\beq
\label{cond4.eq}
\la r\modelur \bigd r\repsrpow \ra =
\modv  r\repsrpow.
\eeq
It follows that the left-hand-side of the Eq.~\ref{cond4.eq} is only
a function of $r\repsrpow$ for $r \ll L$, if the first postulate is valid.
Figure \ref{smallscexp2.fig} shows the logarithm of the 
left-hand-side
of 
\ref{cond4.eq} plotted against the logarithm of $r\repsrpow$ for different
spatial separations in the small $r$ range.
The curves collapse on to one another except possibly at the tails
where statistical uncertainty can be large, indicating that
the first hypothesis is approximately valid at $\rel \sim 650$. 

The results corresponding to the small $r$ regime presented
here
show 
support for the first refined similarity hypothesis. 
In contrast, previous work 
(Ref.~[\onlinecite{krsh}])
on this topic 
has found 
evidence against the first similarity hypothesis.
Subsequently the authors of 
Ref.~[\onlinecite{krsh}]
used an ad-hoc dependence on $r$ and $\epsr$
to empirically determine the functional form of $V$ at the smallest scales. 
We are of the opinion that the uncertainty in the
PDF of $V$ in 
Figs.~$4$(a) and $4$(b) of 
Ref.~[\onlinecite{krsh}] 
is at least partly a consequence of
inadequate sampling.
Effects of limited sampling are especially pronounced
in the small $r$ regime, because the averaging intervals
are small.
The problem of limited sampling is alleviated at least partly 
in our work by the use of 3D local averages of dissipation and the
use of
three different 
samples of $V$ along the three orthogonal directions
(Eq.~\ref{valpha.eq}).
Consequently, there is less noise in the statistics of $V$,
even when the averaging intervals are small ($r \sim \eta$). We have
checked that decreasing the sample size by considering velocity
increments $\delur$ only along one direction leads to considerable
noise in the PDF of $V$ at the  smallest scales.  
\vspace{-5mm}
\section{Conclusions}
We have compared and contrasted three-dimensional (3D) and 
one-dimensional (1D) local averages of dissipation. 
Since turbulence is three-dimensional, 
the 3D averaging is nominally the right procedure while 1D averaging 
has long been practiced as a matter of convenience. 
Further, the 3D averages of dissipation are influenced less 
by factors such as finite sampling and periodic boundary 
conditions than the 1D averaged dissipation. 
In particular, results from 3D averaging are less intermittent 
at the small scales than the 1D case, 
rendering the former more statistically stable than the latter.

Considering the advantages of 3D averaging, 
it seemed appropriate to use 3D averaged dissipation to 
examine the first and second postulates of K62 for a $4096^3$ data set at 
$\rel \sim ∼ 650$. 
The basic tenets of the first and second K62 postulates were found to be 
true. 
The variance and skewness of the variable $V$ in the inertial range 
are shown to be consistent with the corresponding K62 predictions and 
$V$ is approximately universal in this scale range. 
At the smallest scales, the statistics of $V$ seem to depend only on the 
local Reynolds number to a good approximation.

In light of the support for the refined similarity hypotheses 
shown in this work, at least at the level of detail examined here, 
the following observations appear pertinent. 
For any singularity of exponent $\alpha$ of $\repsr$, where $\epsr$
is the 3D local averaged dissipation and $r \ll L$, 
there exists an associated 
singularity of exponent $h = \alpha/3$ for the velocity on the 
same set with fractal dimension $D(h)$, such that
\beq
\label{mfracall.eq}
h = \frac{\alpha}{3}, \quad D(h) = f(\alpha), 
\quad \psip = \frac{p}{3}+\taup \;;
\eeq
where $f(\alpha)$ is the multifractal singularity spectrum 
\cite{Fri95,menevkrs}. We are aware of objections of principle raised in 
\cite{frisch91} with regard to the refined similarity hypotheses 
and also of numerical results of \cite{SSY07} in which an 
alternative theory \cite{Y01} seemed to
agree marginally better with the simulations data. 
We point out that the previous work has not taken account of 
3D averaging, which we advocate here as being more appropriate. 
It remains to be seen whether the refined similarity hypotheses 
stand when a more detailed 
assessment is attempted.

\section{Acknowledgments}
We thank Rich Vuduc for his 
help with the computational aspects of this work.
KI acknowledges  funding from the European Research Council under the European 
Community’s Seventh Framework Program, ERC Grant Agreement No $339032$.
This work is partially supported by the National Science
Foundation (NSF), via Grants CBET-$1139037$, OCI-$0749223$
and OCI-$1036170$ at the Georgia Institute of Technology.
The computations were performed using supercomputing resources
provided through the XSEDE consortium (which is funded by NSF)
at the National Institute of Computational Sciences
at the University of Tennessee (Knoxville), the Texas Advanced
Computing Center at the University of Texas (Austin),
and the Blue Waters Project at the National Center for
Supercomputing Applications at the University of Illinois
(Urbana-Champaign).

\bibliography{zebib}
\end{document}